\documentclass[preprint]{emulateapj}
\usepackage{epsfig} 
\usepackage{epstopdf}
\usepackage{graphicx}
\usepackage{amsmath} 
\usepackage{hyperref}
\usepackage{tabularx}
\usepackage{aas_macros}

\usepackage{breakcites}

\newcolumntype{R}{>{\centering\arraybackslash}X}

\begin{document}

\title{Forbidden Iron Lines and Dust Destruction in Supernova Remnant Shocks: The Case of N49 in the Large Magellanic Cloud}

\author{Michael A. Dopita\altaffilmark{1,2}, Ivo R. Seitenzahl\altaffilmark{1}, Ralph S. Sutherland\altaffilmark{1}, Fr\'ed\'eric P. A. Vogt\altaffilmark{3}, \newline P. Frank Winkler\altaffilmark{4},  \& William P. Blair\altaffilmark{5}}
\email{Michael.Dopita@anu.edu.au}

\altaffiltext{1}{Research School of Astronomy and Astrophysics, Australian National University, Canberra, ACT 2611, Australia.}
\altaffiltext{2}{Astronomy Department, King Abdulaziz University, P.O. Box 80203, Jeddah, Saudi Arabia.}
\altaffiltext{3}{ESO - European Southern Observatory, Av. Alonso de C\'ordova 3107, 7630355 Vitacura Santiago, Chile}
\altaffiltext{4}{Physics Dept., Middlebury College, Middlebury, VT 05753, USA}
\altaffiltext{5}{Department of Physics and Astronomy, Johns Hopkins University, Baltimore, MD 21218, USA}

\begin{abstract}
We present results of a complete integral field survey of the bright SNR N49 in the LMC, obtained with the WiFeS instrument mounted on the ANU 2.3m telescope at  Siding Spring Observatory. From theoretical shock modelling with the new \emph{MAPPINGS 5.1} code we have, for the first time, subjected the optical Fe emission line spectrum of a supernova remnant to a detailed abundance and dynamical analysis covering 8 separate stages of ionisation. This allows us to derive the dust depletion factors as a function of ionisation stage. We have shown that there is substantial (30\% -- 90\%) destruction of Fe-bearing dust grains in these fast shocks ($v_s \sim 250$\,km\,s$^{-1}$), and we have confirmed that the dominant dust destruction is through the non-thermal sputtering and grain-grain collision mechanisms developed in a number of theoretical works.
\end{abstract}

 
\section{Introduction}
Since its identification as a supernova remnant (SNR) on the basis of its strong [\ion{S}{2}] emission relative to H$\alpha$ and its association with a bright non-thermal radio source by \citet{Mathewson63} and \citet{Westerlund66}, N49 in the Large Magellanic Cloud (LMC) has been the subject of many detailed studies. Because it is a relatively young SNR with an age of about 5000 yr \citep{Park12}, and because it is located in the vicinity of a dense, dusty molecular cloud in the interstellar medium (ISM) \citep{Banas97,Otsuka10} it has entered into its radiative phase rather quickly. As a result, it is both unusually luminous at optical wavelengths, and is also showing clear signs that the diffuse interior hot plasma mapped at X-ray wavelengths by \citet{Park12}  is itself starting to cool and recombine \citep{Uchida15}.

With the help of space-based observatories, N49 has been studied in exquisite detail both in the spectral and spatial domain, notably by archival X-ray and HST \citep{Vancura92a, Bilikova07}, HUT \citep{Vancura92b} and FUSE \citep{Sankrit04}. These studies reveal that the SNR seen at X-ray wavelengths is diffuse and filled, while the optical images show a very fine filamentary structure in the optical forbidden lines, with marked differences between the structures seen in [\ion{Ne}{5}],  [\ion{O}{3}], [\ion{S}{2}]  and H$\alpha$. 

The first quantitative optical spectroscopy of N49 was given by \citet{Osterbrock73}, who were also the first to point out the rich forbidden line spectrum of [\ion{Fe}{2}] and  [\ion{Fe}{3}], along with the presence of emission lines from other refractory elements such as Mg I and Ca II. This provided the first indication that dust grains are being destroyed in the radiative shocks. This spectrum was later analysed by \citet{Dopita76}, who derived a first estimate of the chemical abundances and shock conditions based upon radiative shock wave theory.

The coronal line of  [\ion{Fe}{14}] $\lambda5303$ was detected spectroscopically by \citet{Murdin78}, and confirmed by \citet{Danziger85}, proving that gas is being heated in shocks up to about $1.5\times10^6$K. This requires a shock velocity in excess of 300\,km\,s$^{-1}$. Inspired by this discovery, \citet{Dopita79} performed narrow-band imaging in the light of this ion, and found that the  [\ion{Fe}{14}] $\lambda5303$ emission was distributed both over a broader spatial region, and more evenly than in H$\beta$. This implies that the  [\ion{Fe}{14}]  emission arises immediately behind the outer blast wave of the SNR.

Despite the intrinsic potential of the Fe forbidden line spectrum as a diagnostic tool \citep{Osterbrock73}, the absence of the detailed atomic physics parameters needed to compute the Fe spectrum has precluded the use of these lines up to now. However, as we will show, the SNR N49 reveals an enormously rich forbidden line spectrum, with Fe seen in no less than eight separate stages of ionisation.  In this paper, we combine new integral field spectroscopy with vastly improved radiative shock modelling to study kinematic differences between different ions of Fe, and to discover how grain destruction works to release Fe back into the gas phase through the cooling zone of radiative shocks. 

\section{Observations and Data Reduction}
The spectroscopic data were obtained using the Wide-Field Spectrograph \citep[WiFeS][]{Dopita07,Dopita10}. This instrument is mounted at the Nasmyth focus of the Australian National University 2.3m telescope located at the Siding Spring Observatory, Australia.  WiFeS is an image-slicing double-beam integral-field spectrograph.  The spectra were acquired in `binned mode', providing 25$\times$38 spaxels each $1\times$\,1\arcsec\ in angular size. The instrument is a double-beam spectrograph providing independent channels for each of the blue and the red wavelength ranges. We used the B3000 and R7000 gratings, covering the waveband $350-710$~nm in a single observation, and giving a resolution of $R=7000$ in the red ($530-710$~nm), and $R=3000$ in the blue ($340-570$~nm), and providing a velocity resolution of $\Delta v =45\,$km\,s$^{-1}$ in the red and of $\Delta v =100$\,km\,s$^{-1}$ in the blue.

The data on N49 were obtained on the nights of December 18-21, 2014, and consisted of 2x900s exposures on each of 12 overlapping fields. For each field a 900\,s sky reference background frame was also taken and this was subtracted from the data during reduction. Absolute photometric calibration of the data cubes was made using the STIS spectrophotometric standard star HD\,26169\footnote{Fluxes available at: \newline {\url{www.mso.anu.edu.au/~bessell/FTP/Bohlin2013/GO12813.html}}}.  Separate corrections for the OH and H$_2$O telluric absorption features were made. During the observations of the target, the seeing was between 1.5 and 2.5$^{\prime\prime}$, which is reasonably well matched to the 1.0$^{\prime\prime}$ pixels of the spectrograph. Arc and bias frames were also taken regularly, and internal continuum lamp flat fields and twilight sky flats were taken in order to provide sensitivity corrections in both the spectral and spatial directions.

The data were reduced using the {\tt PyWiFeS v0.7.3} pipeline written for the instrument \citep{Childress14a, Childress14b}. In brief, this produces a data cube which has been wavelength calibrated, sensitivity corrected (including telluric corrections), photometrically calibrated, and from which the cosmic ray events have been removed. Because the  {\tt PyWiFeS} pipeline uses an optical model of the spectrograph to provide the wavelength calibration, the wavelength solution is good across the whole field, and does not rely on any interpolation of the data, since each spaxel is assigned a precise wavelength and spatial coordinate. The only interpolation occurs when constructing the final data cube, sampled in wavelength intervals of 0.768\,\AA\ in the blue mosaic and 0.439\,\AA\ in the red mosaic.

The mosaic was assembled using a custom {\tt python} script written by Vogt and available on request. The FITS headers did not originally contain reliable WCS information. These were assigned by anchoring the mosaic to a star in the 2MASS catalogue. The respective alignment of the different WiFeS fields in the mosaic was derived by comparing the reconstructed white-light image from the red cubes with the Digitized Sky Survey (DSS) 2 red band image of the area. Given the mean seeing conditions during the observations ($\sim  2.0^{\prime\prime}$), the spatial shifts are rounded to the nearest integer for simplicity, and also to avoid superfluous resampling of the data. All data cubes are on the same wavelength grid in the data reduction process, so that no shift is required along that axis. The final mosaic has dimensions $ 98^{\prime\prime} \times 91^{\prime\prime}$ on the sky, which for an assumed distance of the LMC of 49\,kpc \citep{Pejcha12} corresponds to a field of $23.5 \times 21.8$\,pc. 

\begin{figure}[htb!]
\begin{centering}
\includegraphics[scale=0.4]{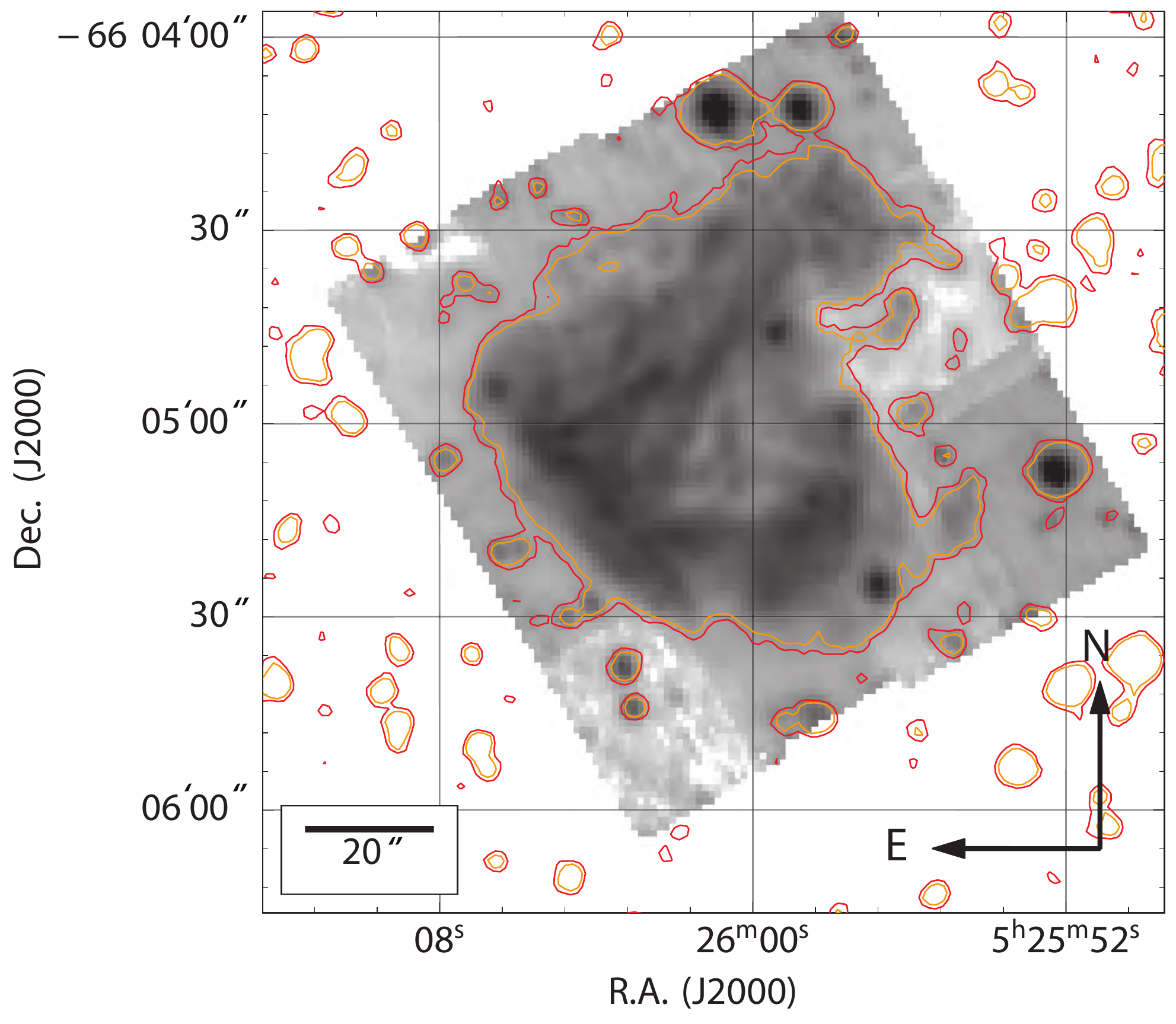}
\end{centering}
\caption{The WiFeS mosaic of N49 (greyscale) compared with the DSS2-R map (orange and yellow contours). In order to be comparable in wavelength coverage to the DSS image, the WiFeS image is a sum of all slices between 5533\AA and 6620\AA. Note that the sky background level in two or three of the outer fields is low. These fields were obtained in non-photometric conditions, and the fluxes are less accurate.} \label{fig1}
\end{figure}

In Figure \ref{fig1}, we present the final alignment of the different WiFeS fields against the DSS-2 red band image of the area.\footnote{The DSS-2 image was obtained from the ESO Online Digitized Sky Survey: \newline {\url{http://archive.eso.org/dss/dss}}}. The final accuracy of the spatial alignment of each individual field is of the order of 1.0\,\arcsec. Since the outermost fields were taken under non-photometric conditions these have a lower S/N ratio.

\section{Images in the Iron Lines}
\subsection{Emission Line Maps}
\begin{figure*}[htb!]
\begin{centering}
\includegraphics[scale=0.90]{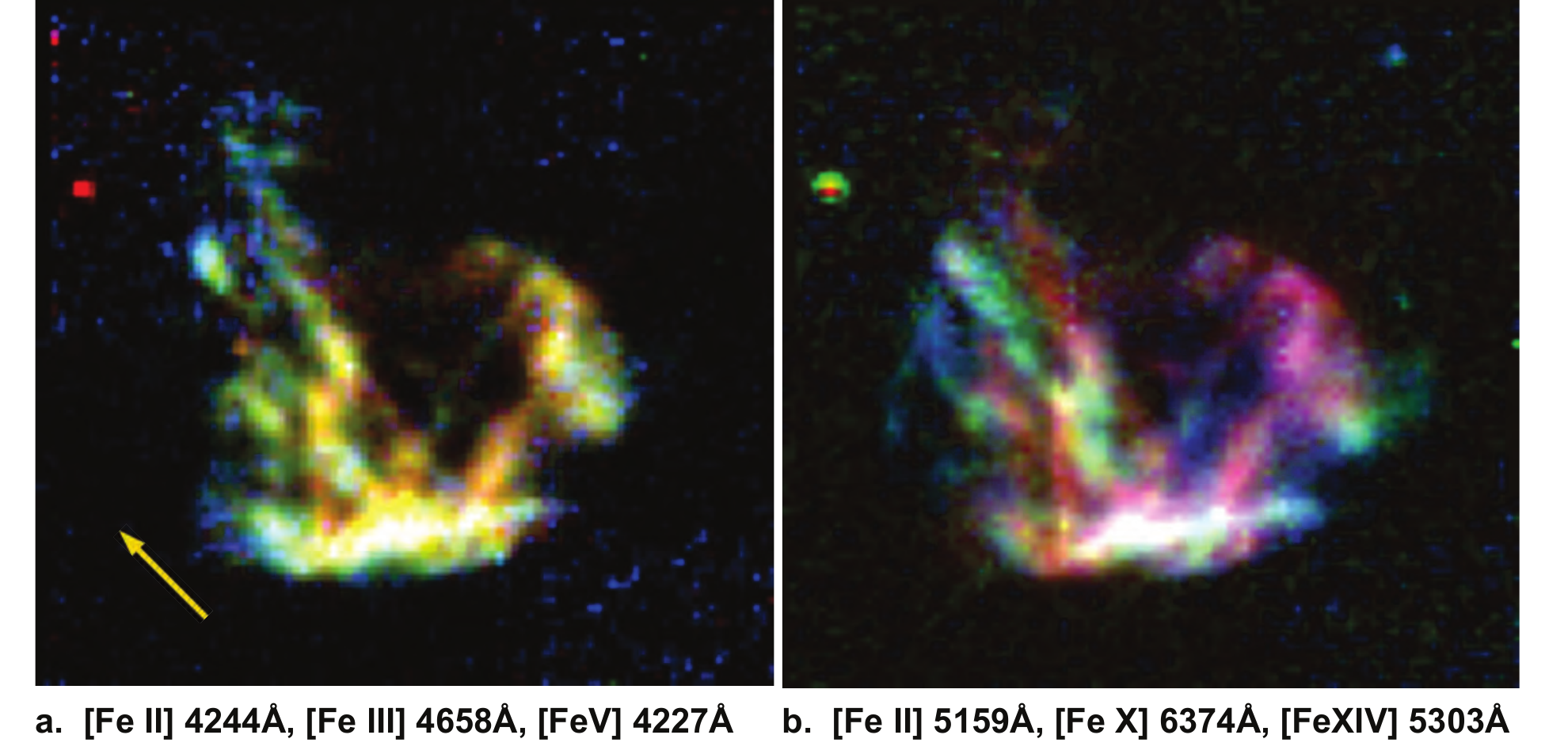}
\end{centering}
\caption{Images of N49 in the forbidden Fe lines. The direction of the yellow arrow indicates North on this and all subsequent images. The images cover a field of $ 98^{\prime\prime} \times$ $91^{\prime\prime}$.The color channels (R,G,B) of each panel are as indicated under each image. There is a clear progression in that the high-excitation lines are much more extended in the NE-SW direction, corresponding to high-velocity shocks seen in projection, while the filaments which are strong in [Fe II] are concentrated closer to the center of the remnant.} \label{fig2}
\end{figure*}

From the mosaics, we have extracted maps of N49 in each of the ionisation states of the Fe forbidden iron lines which are strong enough to be used for this purpose. The results are shown as RGB images in Figure \ref{fig2}. 

There is a clear spatial segregation of the different ionic states. As also argued by \citet{Bilikova07}, this cannot be simply due to the post-shock cooling length being seen in projection. To anticipate the results presented in the detailed modelling (below), a typical [\ion{Fe}{2}]-emitting shock has $v_s = 250$\,km\,s$^{-1}$ and a pre-shock hydrogen density $n_{\rm H} = 80$\,cm$^{-3}$. These numbers imply a cooling length of just 0.04\,pc, much smaller than our $\sim 0.3$\,pc spatial resolution.  However, the cooling timescale is more interesting. For the shock parameters given above, the time taken to cool to a temperature of 5000K is 1500\,yr. 

In order to produce a strong [\ion{Fe}{14}] line, we require a shock velocity of 350-450\,km\,s$^{-1}$. This provides a correspondingly longer timescale to produce a fully radiative shock - 4000 to 5000\,yr. However, if the pressure driving the  [\ion{Fe}{14}]-emitting shocks is similar to that driving the  [\ion{Fe}{2}]-emitting shocks, then the pre-shock density in the  [\ion{Fe}{14}]-emitting shocks  has to be appreciably lower, in the ratio of the square of the shock velocity. This will increase the cooling timescale of these shocks to $\sim 13000$\,yr. Assuming that the outer blast wave is moving at $\sim 400$\,km\,s$^{-1}$, and given that the radius of N49 is $\sim 7$\,pc, the dynamical age given by the Sedov-Taylor theory \citep{Sedov59} is $\sim 8500$\,yr. This value can be compared with the Sedov age of 4800 \,yr derived from deep \emph{Chandra} observations by \citet{Park12}. Either of these estimated ages are appreciably shorter than the timescale for post-shock cooling in the  [\ion{Fe}{14}] emitting shocks. Therefore, it is highly unlikely that these are radiative. This conclusion is also supported by the fact that the  [\ion{Fe}{14}] emission is so spatially distinct. These considerations also apply, but less forcefully, to the  [\ion{Fe}{10}] emission.

The spatial segregation of the different ionisation states supports the conclusion that the phase structure of the ISM, combined with the resulting variation in shock velocity and radiative lifetime, together determine the appearance of the SNR as seen in  the different stages of ionisation of Fe.

\subsection{Velocity Maps}

\begin{figure*}[htb!]
\begin{centering}
\includegraphics[scale=0.88]{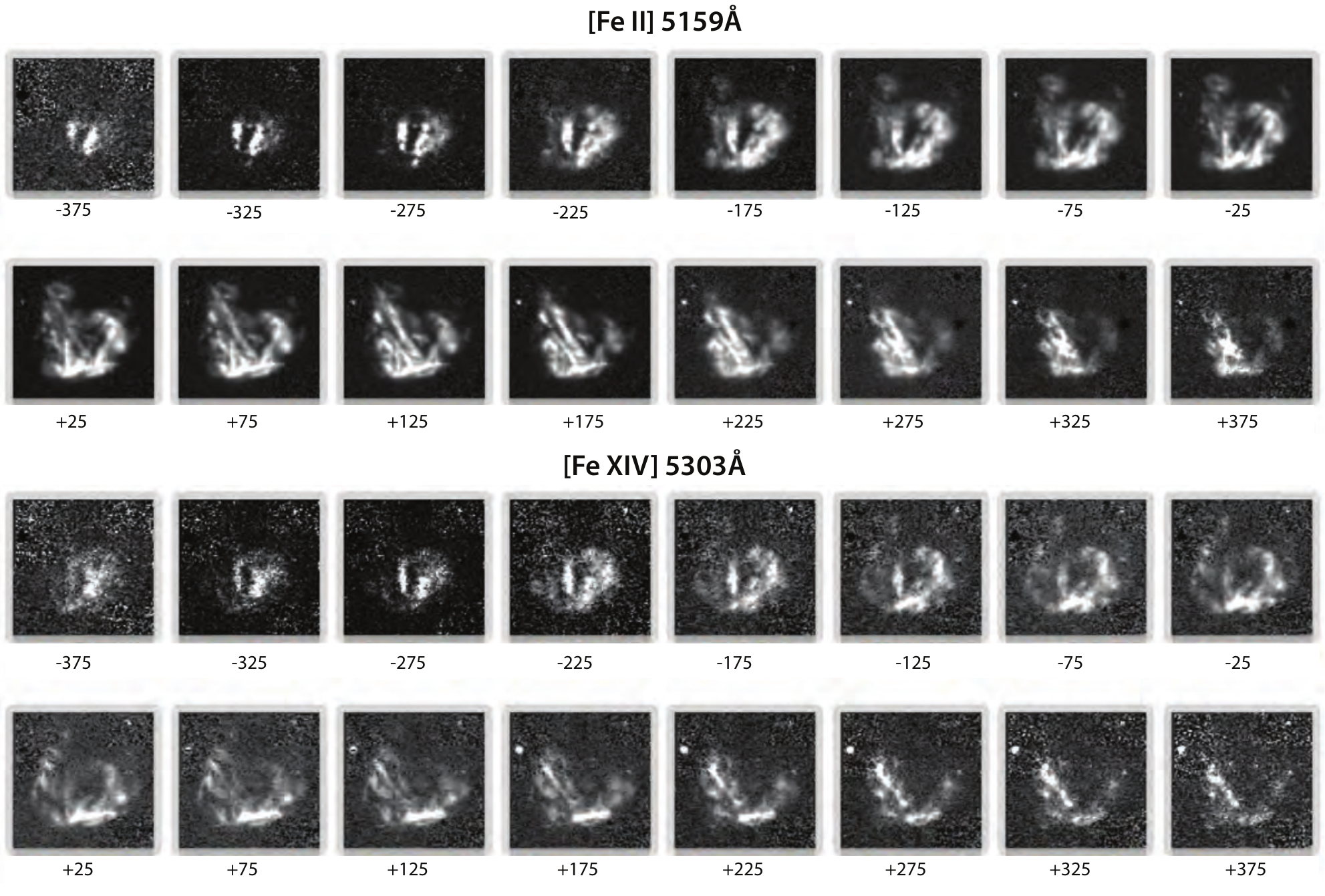}
\end{centering}
\caption{Velocity channel maps for the [\ion{Fe}{2}] $\lambda 5159$ and [\ion{Fe}{14}] $\lambda 5303$ lines. The orientation of the images is the same as indicated in Figure (2). Each channel is marked with its radial velocity with respect to the rest wavelength in the frame of the supernova remnant. Bright or dark dots that appear in many channels are residual star artefacts resulting from the sky subtraction process. The velocity structures indicate that, while some features are correlated, many filaments show little correlation between the two ions, indicating that the [\ion{Fe}{14}] emission arises from fast shocks moving into a lower-density phase medium than those which give rise to the [\ion{Fe}{2}] emission. Note in particular the faint outer blast wave visible in [\ion{Fe}{14}] in the channels -125 through +125 km\,s$^{-1}$.} \label{fig3}
\end{figure*}

The conclusion that the different ionisation stages of Fe arise in different phases of the ISM is also supported by the appearance of N49 in velocity channel maps of N49 of the  [\ion{Fe}{2}] and  [\ion{Fe}{14}] lines (see Figure \ref{fig3}). Strong differences in the morphology appear in the channels around zero velocity, where the full extent of the SNR appears in the  [\ion{Fe}{14}] emission. This supports the general picture advocated by \citet{Shull85}, and which is re-enforced by the multi-wavelength analysis of \citet{Bilikova07}. In this model, four phases are present:
\begin{enumerate}
\item{Optical emission arising from slower shocks moving into dense atomic and molecular clouds}
\item{Hard X-rays arising from the diffuse ejecta and shock-heated ISM}
\item{Soft X-rays (and in our case  the [\ion{Fe}{14}] emission) from the outer blast wave and}
\item{Partially-radiative optical emitting shocks arising in the denser phases of the ISM associated with the outer blast wave -- which in our case appear in the [\ion{Fe}{10}], and to some degree, in the [\ion{Fe}{5}] emissions.}
\end{enumerate}

\subsection{Global Spectrum of N49}
From the data cube, we extracted a global spectrum of N49 from an aperture of $ 64^{\prime\prime}$ diameter centred on the remnant. This spectrum is already sky-subtracted using the sky observations. However, in order to ensure the removal of any residual sky features we subtracted an annular region with $ 70^{\prime\prime}$ outer diameter and $ 66^{\prime\prime}$ inner diameter. This spectrum is presented in Table \ref{Table1}. Here the spectrum has been de-reddened with the \citet{Cardelli89} reddening law using the same E(B-V) = 0.35 (logarithmic reddening constant of 0.5) as \citet{Vancura92a}. The de-reddened line fluxes given in the Table are normalised to H$\beta = 100.0$.

For the line identification, the spectrum was shifted to rest wavelength, and the wavelengths of the line identifications are given from the {\tt Mappings 5.1} code (Sutherland et al. 2015, in prep.){\footnote{Available at {\url http://miocene.anu.edu.au/Mappings}} which uses data from NIST (US Government National Institute of Standards and Technology) public domain Atomic Spectroscopy Database; NIST Standard Reference Database \#78{\footnote {\url http://www.nist.gov/pml/data/asd.cfm}} (2014). This is currently the most comprehensive source of critically evaluated atomic data for atomic ionisation potentials, energy levels and transition data.  The air wavelengths for all transitions are derived using the  \citet{Peck72} atmospheric refractive index data. To facilitate comparison with the shock model, the details of which are presented below, we have identified in Table \ref{Table1} the individual components of blended lines. These are distinguished by the letter (B) in the Table.

The spectrum given in Table \ref{Table1} emphasises the importance of the forbidden Fe emission line spectrum in this object. Iron is identified in the following stages of ionisation: \ion{Fe}{2}, \ion{Fe}{3}, \ion{Fe}{5}, \ion{Fe}{6}, \ion{Fe}{7}, \ion{Fe}{9}, \ion{Fe}{10} and \ion{Fe}{14}. In addition, a number of other refractory elements are identified, Ni in the stage \ion{Ni}{2}, Ca in the stages \ion{Ca}{2} and (possibly) \ion{Ca}{7} and also Mg in the ion  \ion{Mg}{1}. The rich spectrum of refractory elements shows how radiative shock waves can -- in principle -- be used as their own diagnostic tests of grain destruction occurring within them.

\begin{table*}[htbp]
\tiny
\caption{The global emission spectrum of N49 de-reddened by $E(\rm B- V) = 0.35$ using the \citet{Cardelli89} reddening law compared with the single shock model described in the text.}\label{Table1}
\begin{center}
\begin{tabular}{lcrrclcrr}
\\ \hline \hline 
Ion & Rest & Intensity & Model &  & Ion & Rest & Intensity & Model \\
ID & $\lambda$(\AA) &(H$\beta =100$) & &  & ID & $\lambda$(\AA) & (H$\beta =100$) & \\
\hline 
 &  &  &  &  &  &  &  & \\
 {H I} & 3703.9&  $1.55\pm0.25$ & 1.23&   & {[Ar IV]} & 4740.1&  $0.35\pm0.10$ & 0.60 \\
 {H I }& 3711.9&  $1.38\pm0.25$ & 1.48&   & {[Fe III]} & 4754.7&  $0.81\pm0.20$ & 0.53 \\
{[O II] (B)} & 3726.0&  $968\pm18.0$ & 479.00&   &  {[Fe III] } & 4769.4&  $0.35\pm0.15$ &  0.36 \\
{[O II] (B)} & 3728.8&   & 331.00&   &    {[Fe II] } & 4774.7&   $0.52\pm0.15$ &  0.51\\
{H$\kappa$} & 3750.2&  $3.09\pm0.35$ & 2.91&   &  {[Fe II]} & 4798.3&  $0.13\pm0.05$ &  0.15 \\
{[Fe VII]} & 3758.9&   $0.48\pm0.20$ & 0.35&   &  {[Fe II]} & 4814.5&  $2.45\pm0.20$ &  3.38 \\
{H$\iota$} & 3770.6&  $4.24\pm0.25$ & 3.79&   &  {[Fe II]} & 4874.5&  $0.35\pm0.10$ &  0.52 \\
{H$\theta$} & 3797.9&  $5.77\pm0.20$ & 5.07&   &  H$\beta$ & 4861.3&  $100.00\pm1.50$ &  100.00 \\
{He I} & 3820.0&  $1.28\pm0.15$ & .... &   & {[Fe II]} & 4874.5&  $0.35\pm0.10$ &  0.52 \\
 {H$\eta$} & 3835.4&  $8.56\pm0.40$ & 6.99&   &  {[Fe III]} & 4881.0&  $1.51\pm0.50$ &  2.28 \\
 {[Ne III]} & 3868.8&  $43.02\pm0.80$ & 71.00&   &  {[Fe II]} & 4889.6&  $1.67\pm0.20$ &  2.82 \\
 {He I}, H$\zeta$ & 3888.6&  $24.32\pm0.75$ & 19.70&   &  {[Fe II]} & 4905.4&  $0.75\pm0.40$ &  1.01 \\
{Ca II} & 3933.7&  $16.15\pm0.40$ & 1.90&   &  He I & 4921.9&  $1.11\pm0.25$ &  1.14 \\
{[Ne III](B)} & 3967.5 &  $33.15\pm1.15$ & 21.40&   &  {[Ca VII]} & 4939.5&  $0.17\pm0.10$ &  0.02 \\
{Ca II~~~(B)} & 3968.5 &   & 11.70&   &   {[Fe II]} & 4948.0 &  $0.43\pm0.15$ &  0.53 \\
{H$\epsilon$~~~~~~(B)} & 3970.1&   & 15.30&   &  {[O III]} & 4958.9&  $34.43\pm0.95$ &  60.80 \\
{[Ni II]} & 3993.1&  $0.34\pm0.15$ & 0.07&   & {[Fe VI]} & 4972.5&  $0.42\pm0.20$ &  0.30 \\
{[Fe III]} & 4008.3&  $0.44\pm0.15$ & 0.11&   &  {[Fe III] (B)} & 4986.5&  $1.20\pm0.35$ &  0.73 \\
{He I} & 4026.3&  $2.20\pm0.20$ & 2.00&   &  {[FeVII] (B)} & 4988.5&   &  0.14 \\
{[S II] (B)} & 4068.6&  $27.04\pm1.00$ & 21.30&   &   {[O III]} & 5006.8&  $111.70\pm1.5$ &  176.00 \\
{[S II] (B)} & 4076.3&   $8.96\pm0.35$ & 6.78&   &   He I & 5015.7&  $2.80\pm0.35$ &  2.43 \\
{H$\delta$} & 4101.7&  $26.93\pm0.95$ & 24.90&   &  {[Fe II]} & 5043.5&  $0.30\pm0.15$ &  0.35 \\
{[Fe II]} & 4115.6&  $1.70\pm0.35$ &  .... &   &  {[Fe II]} & 5072.4&  $0.11\pm0.05$ &  0.12 \\
{[Fe V]} & 4143.2&  $0.45\pm0.15$ & 0.12&   &  {[Fe III]} & 5084.7&  $0.17\pm0.05$ &  0.06 \\
{[Fe II]} & 4177.2&  $0.48\pm0.15$ & 0.37&   &   {[Fe II]} & 5111.6&  $1.10\pm0.20$ &  1.46 \\
{[Ni II]} & 4201.7&  $0.29\pm0.15$ & 0.41&   & {[Fe VI]} & 5145.8&  $0.19\pm0.10$ &  0.31 \\
{[Fe II]} & 4212.0&  $0.47\pm0.20$ &  .... &   &   {[Fe II]} & 5158.8&  $10.57\pm0.60$ &  12.50 \\
{[Fe V]} & 4227.2&  $0.72\pm0.30$ & 0.59&   &  {[Fe VI]} & 5176.0 &  $0.26\pm0.10$ &  0.35 \\
{[Fe II]} & 4244.8&  $4.97\pm0.35$ & 4.31&   &  {[N I] (B)} & 5198.9&  $5.00\pm0.25$ &  2.94 \\
{C II} & 4267.0&  $0.18\pm0.10$ & 0.14&   &  {[N I] (B)} & 5200.1&   &  4.94 \\
{[Fe II]} & 4276.8&  $1.42\pm0.25$& 1.68&   &  {[Fe II]} & 5220.1&  $0.51\pm0.20$ &  0.58 \\
{[Fe II]} & 4287.4&  $4.70\pm0.80$ &   &   & {[Fe II] } & 5261.6&  $4.27\pm0.25$ &  4.41 \\
{[Fe II]} & 4305.9&  $0.31\pm0.15$ & 0.12&   &  {[Fe III] (B)} & 5270.4 &   &  0.31 \\
{[Fe II]} & 4319.6&  $0.56\pm0.20$ & 0.47&   &  {[Fe II] (B)} & 5273.4&  $3.23\pm0.25$ &  2.69 \\
{[Fe II]} & 4346.8&  $0.42\pm0.20$ & 0.57&   &  {[Fe XIV]} & 5303.3&  $1.04\pm0.20$ &  0.00 \\
{H$\gamma$} & 4340.5&  $49.50\pm1.50$ & 46.40&   &  {[Fe II]~~(B)} & 5333.7&  $1.36\pm0.20$ &  1.34 \\
{[Fe II]~~(B)} & 4355.0&  $3.00\pm0.70$ & 0.89&   &  {[Fe VI] (B)} & 5335.2&   &  1.91 \\
{[Fe IX] (B)} & 4359.1&   & 0.14&   &    {[Fe II]} & 5376.4&  $0.44\pm0.20$ &  0.49 \\
{[O III]} & 4363.2&  $8.32\pm1.25$ & 12.10&   &  He II~~~(B) & 5411.5&  $0.92\pm0.15$ &  0.84 \\
{[Fe II]} & 4372.4&  $0.28\pm0.10$ & 0.30&   &  {[Fe III] (B)}  & 5412.0 &  &  0.84 \\
{[Fe II]} & 4382.7&  $0.30\pm0.10$ & 0.28&   &    {[Fe VI]} & 5425.2&  $0.18\pm0.05$ &  0.15 \\
{He I} & 4387.9&  $0.56\pm0.25$ & 0.53&   &  {[Fe II]} & 5433.2&  $0.49\pm0.10$ &  0.78 \\
{[Fe II]} & 4416.3&  $4.74\pm0.30$ & 4.26&    &   {[Fe VI]} & 5484.8&  $0.16\pm0.05$ &  0.10 \\
{[Fe II]} & 4432.5&  $0.12\pm0.05$ & 0.17&    &  {[Fe II]} & 5497.0 &  $0.21\pm0.10$ &  .... \\
{[Fe II]} & 4452.1&  $1.33\pm0.35$ & ....  &   & {[Cl III]} & 5517.7&  $0.22\pm0.10$ &  1.20 \\
{[Fe II]} & 4457.9&  $1.05\pm0.35$ & 1.23&   &    {[Fe II]} & 5533.2&  $1.70\pm0.20$ &  0.78 \\
{He I} & 4471.5&  $4.49\pm0.45$ & 4.19&   &   {[Fe II]} & 5556.3&  $0.16\pm0.05$ &  0.15 \\
{[Fe II]} & 4492.6&  $0.51\pm0.20$ & 0.54&   &   {[O I]} & 5577.3&  $1.14\pm0.25$ &  0.71 \\
{[Fe II]} & 4514.9&  $0.20\pm0.10$ & 0.33&   &   {[Fe VI]} & 5630.8&  $0.10\pm0.05$ &  0.11 \\
{[Fe II]} & 4528.3&  $0.20\pm0.10$ & 0.14&   &   {[Fe VI]} & 5676.9&  $0.09\pm0.05$ &  0.13 \\
{He II} & 4541.6&  $0.21\pm0.10$ & 0.54&   & {[Fe VII]} & 5720.7&  $0.27\pm0.05$ &  0.13 \\
{Mg I]} & 4562.5&  $2.82\pm0.30$ & 2.86&   & {[N II]} & 5754.6&  $1.24\pm0.25$ &  1.54 \\
{Mg I]} & 4566.8&  $4.17\pm0.30$ & 4.30&   &  He I & 5875.6&  $10.93\pm0.60$ &  11.60 \\
{[Cr II]} & 4581.7&  $1.24\pm0.30$ & ....  &   & {[Fe VII]} & 6087.0 &  $0.45\pm0.15$ &  0.30 \\
{[Fe III]} & 4607.0&  $0.23\pm0.10$ & 0.18&   &  {[O I]} & 6300.3&  $99.64\pm2.60$ &  73.70 \\
{[Ni II]} & 4628.0&  $0.12\pm0.10$ & 0.33&   &   {[S III]} & 6312.1&  $0.93\pm0.20$ &  6.16 \\
{[Fe II]} & 4639.7&  $0.12\pm0.10$ & 0.14&   & {[O I]} & 6363.8&  $33.65\pm1.75$ &  23.60 \\
{[Fe III]} & 4658.1&  $4.55\pm0.30$ & 2.89&   &  {[Fe X]} & 6374.5&  $0.86\pm0.15$ &  0.80 \\
{[Fe III]} & 4667.0&  $0.23\pm0.10$ & 0.10&   &  {[Ni II]} & 6441.3&  $0.31\pm0.05$ &  0.02 \\
{He II} & 4685.7&  $6.63\pm0.40$ & 9.00&   &  {[N II]} & 6548&  $21.95\pm1.00$ &  23.00 \\
{[Fe III]} & 4701.5&  $1.26\pm0.25$ & 1.05&   & H$\alpha$ & 6562.8&  $297.00\pm3.25$ &  297.00 \\
{He I ~~~(B)} & 4713.2&  $1.12 \pm0.25$ & 0.45&   &    {[N II]} & 6584.5&  $67.96\pm2.00$ &  67.70 \\
{[Ne IV] (B)} & 4716.0&   & 0.94&   &  {[Ni II]} & 6666.8&  $0.28\pm0.10$ &  0.57 \\
{[Ne IV]} & 4725.6&  $0.75\pm0.20$ & 1.54&   &   He I & 6678.1&  $2.87\pm0.25$ &  3.39 \\
{[Fe III]} & 4733.9&  $0.40\pm0.15$ & 0.35&   &  {[S II]} & 6716.4&  $113.65\pm2.75$ &  109.00 \\
{[Ar IV]} & 4740.1&  $0.35\pm0.10$ & 0.60&   &  {[S II]} & 6730.8&  $134.28\pm3.00$ &  134.00 \\

  &  &  &  &  &  &  &  & \\
 \hline
\end{tabular}
\end{center}
\end{table*}

\section{Shock Modelling}
\subsection{Description of Shock Grid}
Up to the present, the optical forbidden lines of iron have not been used in a quantitative way in the interpretation of emission from radiative shocks. The aim of our shock modelling presented here is to rectify this situation.

We have used the {\tt Mappings 5.1} code (Sutherland et al. 2015, in prep.)} to construct a grid of radiative shock models appropriate for N49. This code is the latest version of the {\tt Mappings 4.0} code described in \citep{Dopita13}, and includes numerous upgrades to both the input atomic physics and the methods of solution. In these models we have adopted a chemical abundance set of 0.5 times the local Galactic concordance (LGC) abundances (see Nicholls, 2016, in press). These are  based upon the \citet{Nieva12} observations of early B-star data, and have the advantage that they sample the abundances in the local region of the galaxy (out to 500 pc), providing the current abundances in this region. The  \citet{Nieva12} data also provide the abundances of the main coolants, H, He, C, N. O, Ne, Mg, Si and Fe. For the light elements we use the \citet{Lodders09} abundances, while for all other elements the abundances are based upon \citet{Scott15a, Scott15b} and \citet{Grevesse15}. Thus, in the LGC scale, the ``local region'' reference abundance has $12+\log {\rm (O/H)} = 8.77$, as opposed to the \citet{Grevesse10} solar value of $12+\log {\rm (O/H)} = 8.69$. 

The depletion factors of the heavy elements caused by the condensation of these elements onto dust are defined as the ratio of the gas phase abundance to the total element abundance. The  depletion factors are derived from the formulae of \citet{Jenkins09}, extended to the other elements on the basis of their condensation temperatures and/or their position on the periodic table. In our shock models, we have investigated the effect of changing the logarithmic Fe depletion,  $\log D_{\rm Fe}$ in the range $0.0 > \log \left [ D_{\rm Fe} \right ] > -1.5$.

The actual abundance set used in the models is given in Table \ref{Table2}. In the second column we give the total element abundance, X,  relative to H on the scale $12+\log \left[ \mathrm X/ \mathrm H \right]$. In column (3) we give the logarithmic depletion factor of that element, $\log\left [ D \right ]$ for the particular case of $\log \left [ D_{\rm Fe} \right ] = 1.0$, while in column (4) we give the gas-phase abundance relative to hydrogen; $12+\log \left[ \mathrm {Gas}/ \mathrm H \right]$.

We ran a grid of models adopting  abundances of 1.0 and 0.5 LGC, each with self-consistent pre-ionisation, and with shock velocities of $v_s = 100, 120, 140, 160, 180, 200, 225, 250,$ $275, 300, 350$ and $400$ km\,s$^{-1}$. For each of these velocities we ran a separate grid adopting a transverse pre-shock ratio of gas to magnetic pressure $\beta = 0.5, 1.0$ and 2.0. However, this parameter has little effect on the output spectrum, but simply limits the compression factor achieved in the shock. In what follows we report only on the  $\beta = 0.5$ case. The pre-shock hydrogen density in the grid was taken to be 10cm$^{-3}$. For the specific modelling of N49, we ran a full grid with a higher pre-shock density (80cm$^{-3}$), necessary to reproduce the observed [\ion{S}{2}] $\lambda\lambda 6731/6171$ ratio. For this object we adopt an abundance of $0.5\times$ LGC, with an adjustment in the N abundance in order to better fit the  [\ion{N}{2}]  line intensities. The N abundance used was $12+\log[{\rm N}/{\rm H}]=7.08$.

\subsection{Treatment of the Iron Ions}
Since the analysis presented here depends critically upon the forbidden Fe-line diagnostics, we should explain in more detail here the details of of forbidden Fe-line computations.  Most of the various ions are treated as full multi-level atoms with temperature-dependent collision strengths and the following number of levels in each ion; Fe\,II 107, Fe\,III 108, Fe\,IV 37, Fe\,V 34, Fe\,VI 96, Fe\,VII 9, Fe\,X 3 and Fe\,XIV 5. 

For all ions up to and including FeVII, we use a full multi-level state population calculation which allows for collisional de-excitation effects both to the ground terms and between excited states. This permits the formation of density-sensitive line ratio diagnostics. 

In addition, in the case of Fe X (473 levels) and Fe XIV (315 levels), a multi-level atom is solved for a subset  of levels near ground, and those solutions provide the collision target populations  for collisions to, and subsequent cascades from, the upper levels.  These upper levels are treated in the nebular approximation. That is to say, the ground levels are allowed to be excited into these levels without regard to collisional de-excitation, and the upper levels are then allowed to radiative decay to lower levels, and freely cascade through  intermediate levels back down to the ground term. Numerical tests show that this treatment is valid for electron densities of up to $n_e \sim 10^{12}$\,cm$^{-3}$ and to temperatures beyond $10^7$\,K. The full details will be given in Sutherland et al (in preparation).

Much of the atomic data is derived from the data sources listed in the {\tt Chianti 8.0.3} database  and further data given by \citet{Dere07}, Badnell (2011)\footnote{{\url{http://amdpp.phys.strath.ac.uk/tamoc/DATA/RR/}}} and from the NIST database (Kramida, A., Ralchenko, Yu., Reader, J., and the NIST ASD Team (2012)\footnote{{\url{http://physics.nist.gov/asd3}}}. In summary, the original data sources include (for each ion) \ion{Fe}{2}  \citep{Fuhr88,Nahar95,Zhang95};  \ion{Fe}{3} \citep{Badnell14, Ekberg93}; \ion{Fe}{4} \citep{Froese08}; \ion{Fe}{5} \citep{Nahar00};  \ion{Fe}{6} \citep{Ballance08};  \ion{Fe}{7} \citep{Berrington00} (with corrections by Berrington 2004, private communication); \ion{Fe}{10} \citep{DelZanna04, DelZanna12a, DelZanna12b}. Charge exchange data is derived from  \citet{Kingdon96} and \citet{Ferland97}.

\begin{table}[htbp]
\small
\caption{The abundance set used in the models \newline (for 1.0\,dex depletion of Fe).}\label{Table2}
\begin{center}
\begin{tabular}{lccc}
\\ \hline \hline 
& (2)  & (3) & (4) \\
 Element &	12+[X/H] &	$\log[D]$	& 12+[Gas/H] \\
\hline 
  H  & 12.00 &  -0.00 & 12.00\\
  He & 10.92 &  -0.00 & 10.92\\
  C  & 7.95  & -0.12 &  7.83\\
  N  & 7.10  &  -0.00 &  7.10\\
  O  & 8.46  & -0.02 &  8.44\\
  Ne & 7.79  &  -0.00 &  7.79\\
  Na & 5.91  & -0.18 &  5.73\\
  Mg & 7.26  & -0.31 &  6.95\\
  Al & 6.13  & -0.31 &  5.82\\
  Si & 7.20  & -0.27 &  6.93\\
  S  & 7.05  & -0.00 &  7.05\\
  Cl & 5.20  &  -0.00 &  5.20\\
  Ar & 6.10  &  -0.00 &  6.10\\
  Ca & 6.02  & -1.15 &  4.87\\
  Fe & 7.22  & -1.00 &  6.22\\
  Ni & 5.90  & -0.99 &  4.91\\
  \hline
\end{tabular}
\end{center}
\label{abundances}
\end{table}

\section{Shock Diagnostics}
\subsection{Shock Velocity}
\begin{figure}[htb!]
\begin{centering}
\includegraphics[scale=0.45]{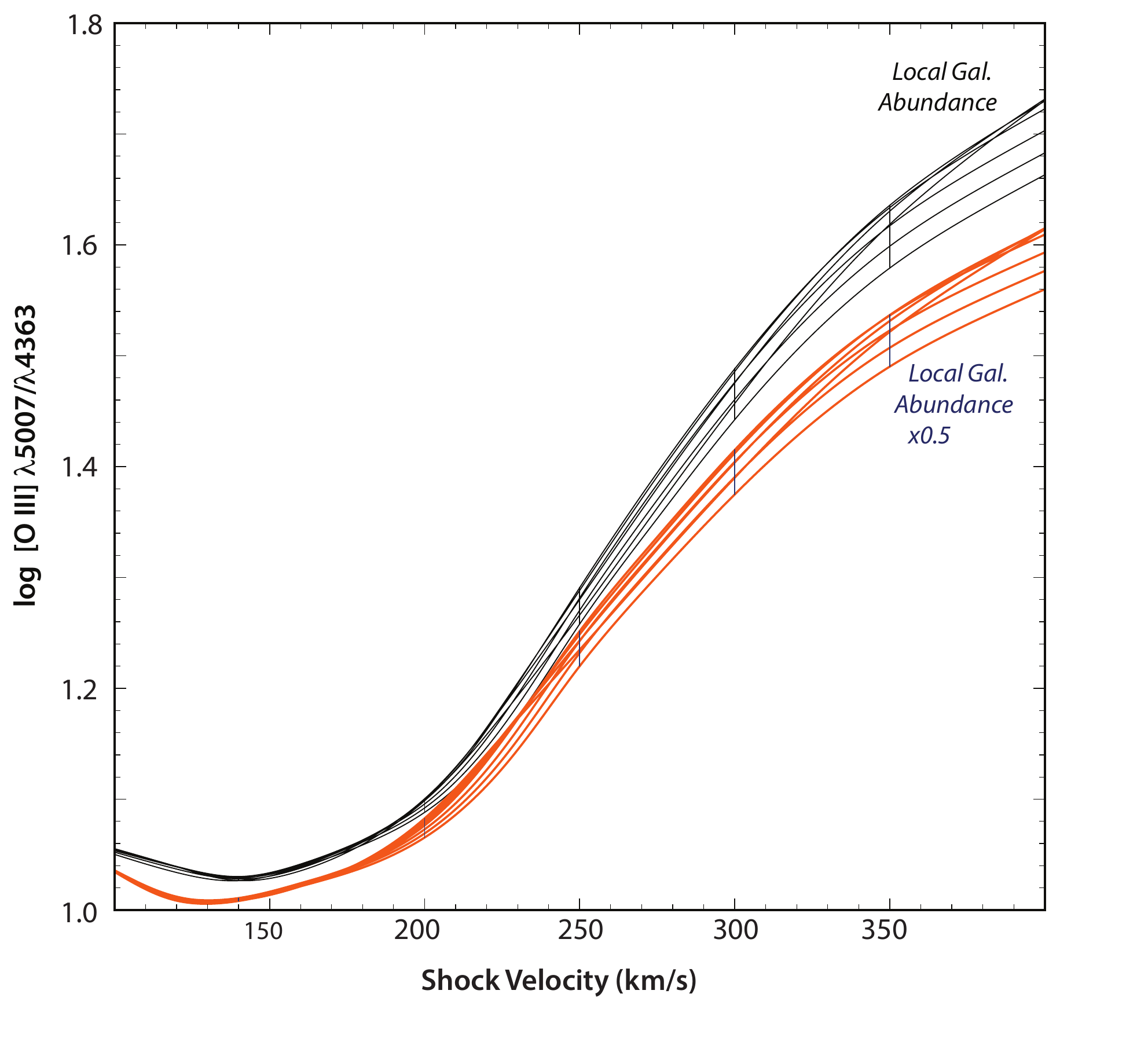}
\end{centering}
\caption{The dependence of  the  [\ion{O}{3}]$\lambda\lambda5007/4363$ ratio on shock velocity and Fe depletion (the unlabelled separate curves). For fully radiative shocks it is evident that this ratio is a fairly good indicator of the shock velocity, with the caveats given in the text.} \label{fig4}
\end{figure}

From the computed grids, we searched for line ratios (not involving the Fe ions) which are strongly dependent on shock velocity, which are sensitive over the full velocity range, and which are not strongly affected by the depletion of Fe. At optical wavelengths we only found one -- the [\ion{O}{3}]$\lambda\lambda5007/4363$ ratio, often used as a temperature diagnostic. The reason that this ratio works as a diagnostic is because the [\ion{O}{3}]-emitting zone of the cooling post-shock gas in low velocity shocks is hot, since the  [\ion{O}{3}] lines arise directly in the cooling zone. However, as the shock velocity increases, the layer of (relatively cool) plasma in the recombination zone of the shock which is photo-ionised by the EUV photons produced in the cooling zone gradually becomes more extensive, and the  temperature indicated by the [\ion{O}{3}]$\lambda\lambda5007/4363$ ratio falls. As the shock velocity increases still further, X-ray Auger ionisation in this zone becomes relatively more important, leading to the development of an extensive region of partly-ionised gas with still lower temperature. This behaviour is evident in Figure \ref{fig4}. Here we show the models for each abundance set with their different Fe depletion factors to illustrate that the  [\ion{O}{3}]$\lambda\lambda5007/4363$ ratio is only very weakly dependent of this parameter.

This diagnostic will fail if the recombining plasma cannot intercept the downstream EUV photons efficiently. This depends on geometry. The model grid assumes a plane-parallel shock structure with efficient EUV photon capture. However, if the shock fronts have high curvature, or a filamentary structure (which seems to be the case in N49 \citep{Vancura92a, Bilikova07,Rakowski07}, then the EUV photons are not captured in the recombination zone, and the [\ion{O}{3}]$\lambda\lambda5007/4363$ ratio will remain much lower, even for fast radiative shocks.

\subsection{Fe-Depletion Diagnostics}
The intensities of the iron lines in the radiative shock spectrum scale rather well with the gas-phase abundance of Fe. The dependence of the forbidden Fe lines on shock velocity is rather weak. This is illustrated in Figure \ref{fig6} for the case of the 
[\ion{Fe}{3}]$\lambda4658/H\beta$ ratio, and in Figure \ref{fig7} for the [\ion{Fe}{2}]$\lambda5156/H\beta$ ratio. For a given shock velocity and abundance set, the [\ion{Fe}{3}]$\lambda4658/H\beta$ ratio scales almost precisely with the gas-phase abundance, which makes this an excellent diagnostic of the gas-phase iron abundance.

\begin{figure}[htb!]
\begin{centering}
\includegraphics[scale=0.42]{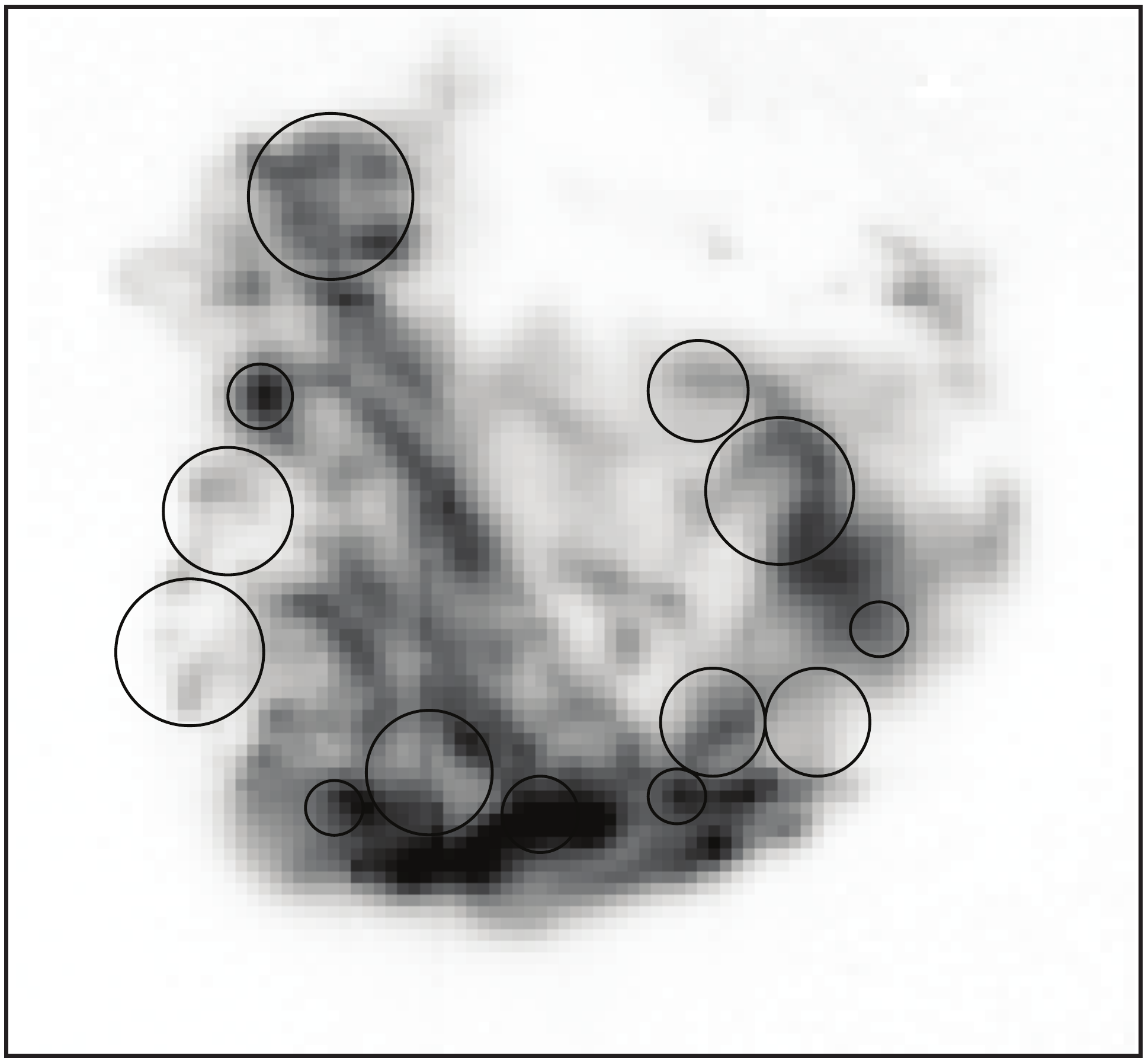}
\end{centering}
\caption{The individual regions in N49 for which the spectra were extracted. The background is an [\ion{O}{3}] $\lambda 5007$\AA\ image.} \label{fig5}
\end{figure}

\begin{figure}[htb!]
\begin{centering}
\includegraphics[scale=0.48]{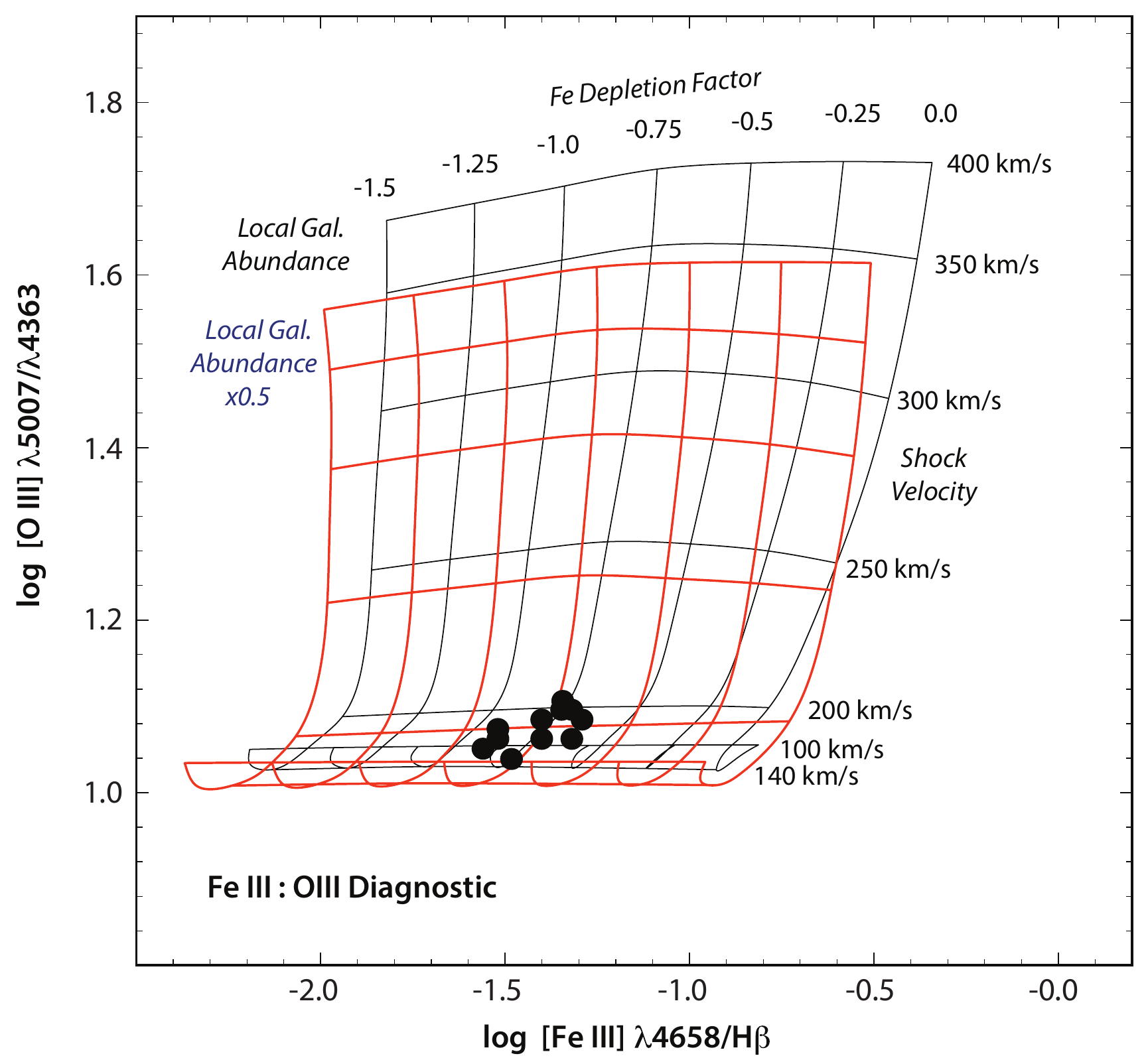}
\end{centering}
\caption{The dependence of the  [\ion{Fe}{3}]$\lambda4658/H\beta$ ratio on shock velocity and Fe depletion. Two grids are shown, for LGC = 1.0 abundance (black) and for LGC = 0.5 abundance (red), which corresponds more closely to the LMC abundance set. The pre-shock density in these models is set at 10\,cm$^{-3}$, and magnetic $\beta = 1.0$. For comparison we show the measured line ratios (black circles) for the circular apertures selected in different regions in N49, see Figure \ref{fig5}, chosen to have different intensities of  the [\ion{Fe}{10}] and [\ion{Fe}{14}] relative to [\ion{Fe}{2}] and [\ion{Fe}{3}]. It is clear that the spectra in  [\ion{Fe}{3}] and [\ion{O}{3}] show little spatial variation, and that the depletion of Fe implied by this [\ion{Fe}{3}] diagnostic is about $\log D_{\rm FeIII} \sim -0.75$.} \label{fig6}
\end{figure}

\begin{figure}[htb!]
\begin{centering}
\includegraphics[scale=0.48]{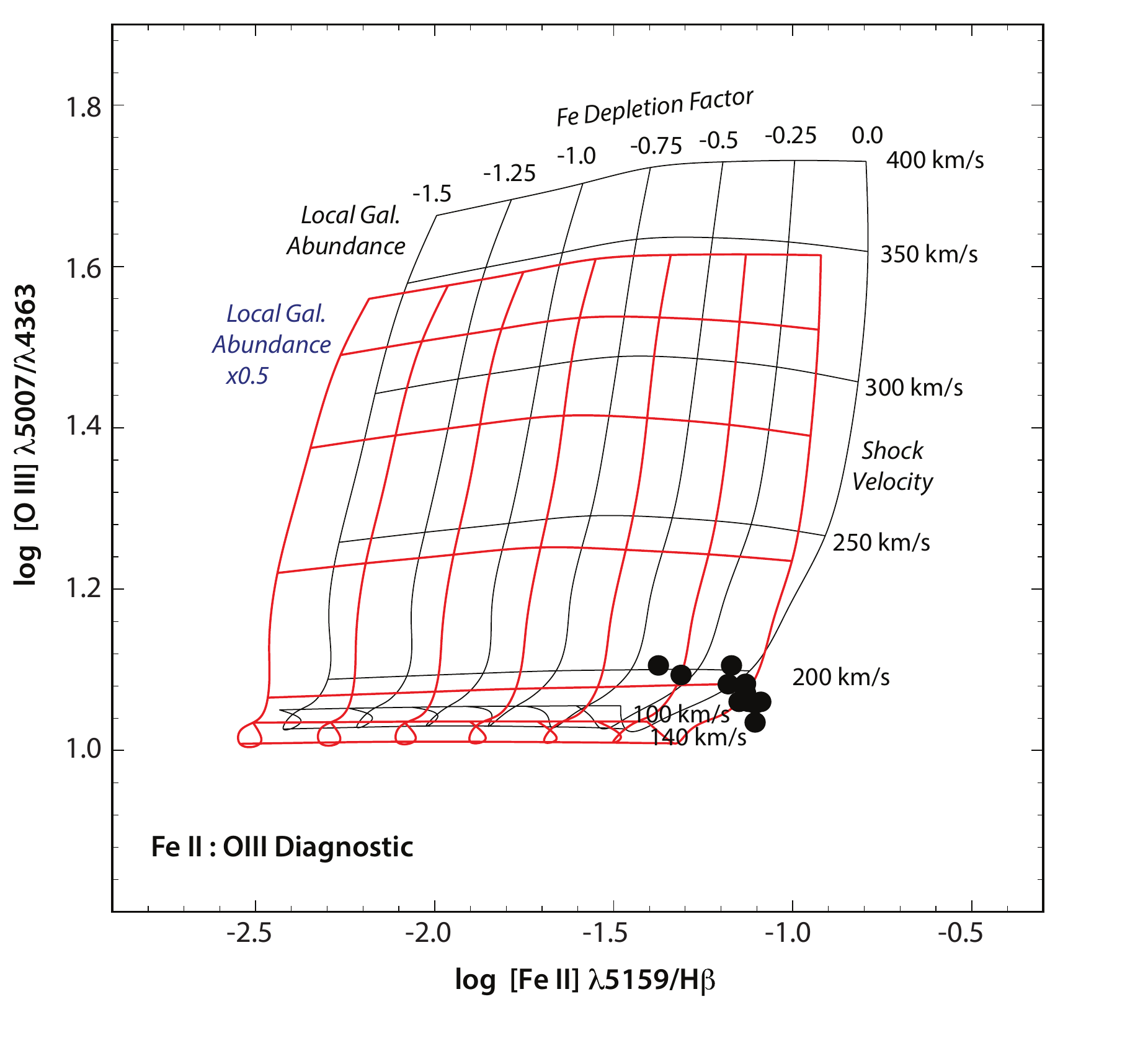}
\end{centering}
\caption{The dependence of the  [\ion{Fe}{3}]$\lambda4658/H\beta$ ratio on shock velocity and Fe depletion. The symbols and regions of N49 represented are the same as in Figure \ref{fig6}. The unexpected result of this diagnostic is that the implied depletion of  this [\ion{Fe}{2}] diagnostic is about $\log D_{\rm FeII} \sim 0.0$, which appears to demonstrate that the Fe-containing dust is only partially destroyed in the [\ion{Fe}{3}]-zone, but is almost fully destroyed by the time the recombination has proceeded to the [\ion{Fe}{2}]-zone.} \label{fig7}
\end{figure}

In order to compare with the observations of N49, we selected a number of regions in the SNR which were characterised by striking differences in excitation as measured by the [\ion{Fe}{10}] or [\ion{Fe}{14}] line intensity relative to [\ion{Fe}{2}] or [\ion{Fe}{3}]. This ought to select between regions of radically different shock velocity, since the  [\ion{Fe}{10}] line only becomes prominent for shock velocities $v_s > 250$\,km\,s$^{-1}$, and the  [\ion{Fe}{14}] line becomes relatively strong for shock velocities in excess of $v_s \sim 350$\,km\,s$^{-1}$. The selected regions are depicted in Figure \ref{fig5}. Mostly they are concentrated around the periphery to avoid regions of line splitting.

A curious result is that the spectra of \emph{all} these sub-regions are very similar, despite having been selected for different excitation conditions (see black circles in Figure \ref{fig6}). This indicates that the shocks which excite the lower-ionisation species of Fe arise in a separate phase than those which give rise to the [\ion{Fe}{10}] or [\ion{Fe}{14}]  emissions. This confirms what has already been adduced from the images and from the dynamics, above.

From Figure \ref{fig7}, the estimated shock velocity in the phase which gives rise to the  [\ion{Fe}{3}] emission is $150 \lesssim v_s \lesssim 200$\,km\,s$^{-1}$. However, this probably represents an under-estimate caused by the filamentary nature of the radiative shocks, as discussed in the previous sub-section. On the basis of the measured velocity dispersions and the expansion velocity of the SNR as a whole, see Figure \ref{fig3}, a characteristic shock velocity of $v_s \sim 250$\,km\,s$^{-1}$ is probably nearer the mark.

A second interesting result is that the depletion of Fe implied by the [\ion{Fe}{3}] diagnostic of Figure \ref{fig6} is about $\log D_{\rm FeIII} \sim -0.75$, indicating that a fairly small fraction of the Fe-bearing grains have been destroyed throughout the post-shock cooling region. This is somewhat surprising, given how hostile this region is expected to be for grain survival against grain-grain collisions and thermal sputtering.

This result is in stark contrast to the [\ion{Fe}{2}] diagnostic of Figure \ref{fig7}. Here, the results are entirely consistent with full dust destruction; $\log D_{\rm FeII} \sim 0.0$. The implication is that the dust destruction rate increases very sharply in the dense recombining gas found on trailing part of the radiative shocks. We now proceed to investigate this result further, using the global spectrum extracted above, see Table \ref{Table1}. 

\section{Global Spectrum Analysis}
\begin{figure*}[htb!]
\begin{centering}
\includegraphics[scale=0.65]{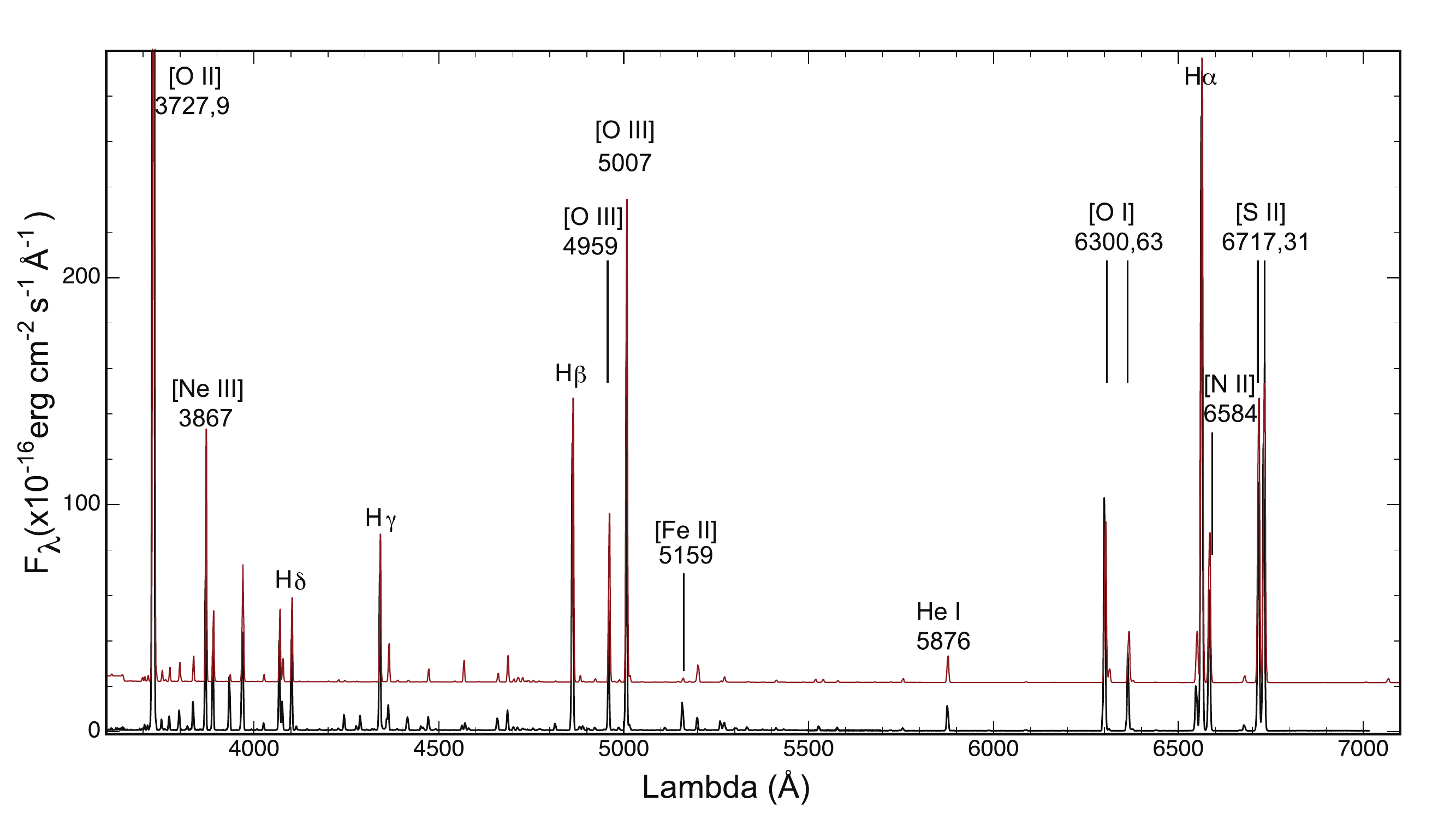}
\end{centering}
\caption{The global (de-reddened) spectrum of N49 (black) fitted by a shock model (red, offset vertically by 20 flux units) with a global abundance of 0.5 times that of the local solar neighbourhood, a pre-shock hydrogen number density of $n_{\rm H}=80\,$cm$^{-3}$, a shock velocity $v_s = 250$\,km\,s$^{-1}$ and a $ \log_{10}$ depletion factor in Fe of -1.0. The theoretical spectrum has been convolved by a Gaussian of FWHM $= 375$\,km\,s$^{-1}$ to fit the observed line width, and has also been scaled to best fit the observed H$\beta$ line profile. This fit is the same as the red theoretical spectrum shown in Figure \ref{fig9}, below. Note that the overall fit is good, but that the [\ion{Fe}{2}] lines such as the $\lambda 5159$ line are predicted to be much weaker in the model, indicating that for this ion, the depletion factor is relatively small.} \label{fig8}
\end{figure*}

In order to investigate Fe in all its ionisation stages, we constructed a series of shock models designed to match the global spectrum of N49 presented in Table \ref{Table1}, transformed to rest wavelengths. The models were designed to give a good fit to the strong emission lines, which indicated a mean shock velocity in the range $200 \lesssim v_s \lesssim 250$\,km\,s$^{-1}$, and a pre-shock hydrogen number density of $n_{\rm H}=80$\,cm$^{-3}$. The theoretical spectrum was convolved with a Gaussian of FWHM $= 375$\,km\,s$^{-1}$ to fit the observed line width, and has also been scaled to best fit the observed H$\beta$ line profile. The result for one of the models is shown in Figure \ref{fig8}. Overall the fit is surprisingly good, given that we are attempting to fit the whole SNR with only a single shock velocity. The predicted line intensities of the ``best-fit'' model are listed in Table \ref{Table3} for the case of $\log D_{\rm Fe} = -1.0$. However, in order to more accurately reflect the observed depletion factors, for the  [\ion{Fe}{2}] lines (only) we give the predicted line intensities for the case $\log D_{\rm Fe} = 0.0$. 

\begin{figure}[htb!]
\begin{centering}
\includegraphics[scale=0.65]{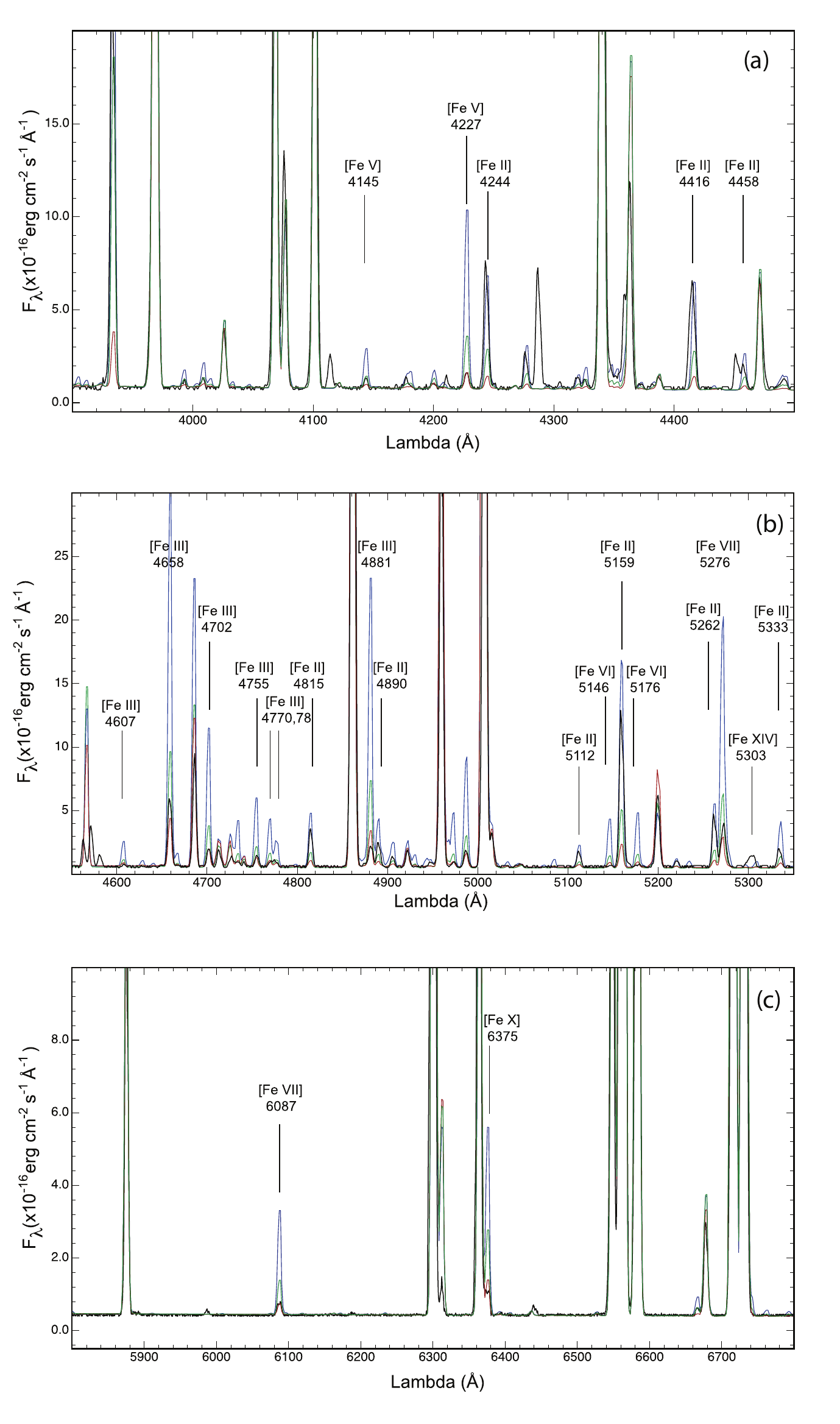}
\end{centering}
\caption{\small The principal forbidden Fe lines used in the depletion abundance analysis.The panels show the observations of the integrated spectrum of N49 (black) overlaid by the scaled theoretical models. The blue curve is for a logarithmic Fe depletion of 0.0, green corresponds to a logarithmic Fe depletion of -0.5, and red to a logarithmic Fe depletion of -1.0. Note that the [Fe II] lines are well fit by a depletion of $\sim -0.2$ while the other Fe lines are all fit with a depletion closer to -1.0.} \label{fig9}
\end{figure}

\begin{table}[htbp]
\small
\caption{The measured depletion factors for Fe.}\label{Table3}
\begin{center}
\begin{tabular}{lccc}
\\ \hline \hline 
 &  &  & \\
Ion & $\left< \log D\right >$ & $\lambda$ & $\log D$ \\
\hline 
 {[\ion{Fe}{2}]} & $-0.16\pm0.07$ & 4177 & -0.25\\
 &  & 4244 & 0.05\\
 &  & 4276 & -0.05\\
 &  & 4416 & 0.00\\
 &  & 4458 & -0.25\\
 &  & 4815 & -0.15\\
 &  & 4890 & -0.20\\
 &  & 5112 & -0.20\\
 &  & 5159 & -0.14\\
 &  & 5262 & -0.10\\
 &  & 5376 & -0.20\\
 &  & 5433 & -0.25\\
 &  &  & \\
{[\ion{Fe}{3}]}& $-0.95\pm0.15$ & 4607 & -1.10\\
 &  & 4658 & -0.80\\
 &  & 4702 & -1.00\\
 &  & 4755 & -0.90\\
 &  & 4769 & -0.75\\
 &  & 4881 & -1.25\\
 &  & 5270 & -1.10\\
 &  & 5412 & -1.00\\
 &  &  & \\
{[\ion{Fe}{5}]} & $-0.90\pm0.20$ & 4143 & -0.60\\
 &  & 4227 & -1.05\\
 &  &  & \\
{[\ion{Fe}{6}]} & $-1.40\pm0.35$ & 5146 & -1.40\\
 &   & 5176 & -1.40\\
 &  & 5425 & -1.60\\
 &  & 5484 & -1.30\\
 &  & 5676 & -1.30\\
 &  &  & \\
{[\ion{Fe}{7}]} & $-0.97\pm0.3$ & 5720 & -1.00\\
 &  & 6087 & -0.95\\
 &  &  & \\
{[\ion{Fe}{10}]} & (-1.25) & 6734 & -1.25\\
  \hline
\end{tabular}
\end{center}
\end{table}

The fit for the depletion factors determined from the most important forbidden Fe lines is shown in Figure \ref{fig9}, and the measured depletion factors determined from fitting to individual lines are listed in Table \ref{Table3}. For the average depletion factor in each ionisation stage we have weighted the data in proportion to the strength of the individual emission lines, since the depletion factors determined from the strongest lines are likely to be the most accurate. For the [\ion{Fe}{10}] species we give the formal fit, but we do not believe this to be accurate, since the strength of the  [\ion{Fe}{10}]  line depends critically on the shock velocity used in the model, and from the discussion above we believe this species to arise in a faster and only partially-radiative component. For the  [\ion{Fe}{14}] line the same considerations apply; we would require a shock with velocity $v_s \sim 375-450$\,km\,s$^{-1}$ to reproduce the observed strength of this line.

In summary, we find -- in agreement with the results in the previous section --- that the best fit to the  [\ion{Fe}{3}] lines, and to lines of higher excitation (with the exception of  [\ion{Fe}{10}]  and [\ion{Fe}{14}]) is consistent with a depletion factor $\log D_{\rm Fe} \sim -1.0$, while the  [\ion{Fe}{2}] lines are consistent with a depletion factor  of only $\log D_{\rm Fe} \sim -0.16$. Thus it is clear that in N49, the Fe-containing grains survive almost unscathed throughout the cooling zones of the blast-wave shocks but are rapidly destroyed in the recombination zone of these same shocks. We will now examine how secure this result is against the uncertainties in the atomic data for Fe\,II, against variations in the gas to magnetic pressure ratio, $\beta$, against thermally unstable cooling, and against varying pre-shock ISM density.

\subsection{Effect of Fe II Atomic Data}
The  \ion{Fe}{2} depletion factors derived above depend critically upon the accuracy of the atomic data sets we have used. The \ion{Fe}{2} ion is notoriously complex, having a multitude of low-lying levels, many of which have critical densities in the range of those frequently encountered in nebulae. There are many radiative cascade paths with comparable importance. In addition both the collision strengths and the radiative transition probabilities have uncomfortable large computational errors. For example, \citet{Bautista15} compare 9 calculations and find a considerable spread. The \citet{Zhang95} collision strengths which we use are, on average,  $0.3-0.5$\,dex weaker than the  \citet{Bautista15} values. On the other hand, the \citet{Ramsbottom07} data lies close to the \citet{Zhang95} values. The critical densities also vary by similar amounts, between  \citet{Bautista15}  and \citet{Zhang95} and \citet{Ramsbottom07}.

A problem would arise were we to adopt the  \citet{Bautista15} collision data as we would predict the \ion{Fe}{2} lines to be weaker by a factor of as much as three.This implies that using these collision strengths would yield much greater ionic abundances compared to those obtained by using those of \citet{Zhang95}, which we employ. Indeed the   \citet{Bautista15} collision data would then require an enhancement in the absolute abundance of Fe, rather than depletion in the \ion{Fe}{2} zone. More than complete grain destruction in the \ion{Fe}{2} zone seems rather unphysical. Nonetheless, our conclusion that the grains are preferentially destroyed in the  \ion{Fe}{2} zone of the shock would remain unaltered.

At face value, therefore, individual \ion{Fe}{2} emission lines may have 0.3-0.5\,dex computational uncertainties. However, when comparing the  line by line depletions derived from the in the N49 spectrum in Table  \ref{Table3}, we find a much smaller spread, and a mean scatter of only 0.07\,dex, which reveals  good consistency between the \citet{Zhang95} data and the observations.

A second factor which may influence the global strength of the  \ion{Fe}{2} emission lines is the ionisation balance. This depends critically on the charge exchange reaction rate between  \ion{Fe}{2} and  \ion{Fe}{1}. In our use of the \citet{Kingdon99} charge exchange rates,   \ion{Fe}{2} is made more abundant in the cooling zone of the shock and consequently the  \ion{Fe}{2} lines are predicted to be stronger by about $\sim0.5$\,dex than when the charge exchange is switched off.  Thus, without charge exchange, the implied depletion factors for  \ion{Fe}{2} will become even smaller -- perhaps even zero. However, the most likely reason that the \ion{Fe}{1} lines are absent from our spectra is probably not due to charge transfer effects, but to the fact that, due to the low ionisation potential of \ion{Fe}{1} (7.9\,eV),  the background stellar UV radiation field is of itself capable of maintaining the ionisation state at the level of \ion{Fe}{2}.  

In conclusion, neither charge exchange reactions nor the newer atomic computations of the  \ion{Fe}{2} ion will alter our conclusion that the grains are preferentially destroyed in the  \ion{Fe}{2} zone of the shock.

\subsection{Effect of Shock Precursor Emission}
The depletion factors that we have determined could be influenced by the effect of shock precursor emission, which does not necessarily have the same ratio of iron forbidden lines to the hydrogen recombination lines. We do know from the work of \citet{Shull83}, and the later work of \citet{Vancura92a} that N49 displays narrow emission lines, particularly in [O\,III] $\lambda\lambda4959,5007$, which are understood to arise from the photoionised precursor of the fast shocks moving into the surrounding interstellar medium \citep{Dopita95,Dopita96}. Although this precursor emission appears to be prominent on echelle spectra, this is simply the result of the low velocity dispersion in the precursor.  In fact, our computations show that the precursor region has low emissivity compared with the main radiative shock, and can at best contribute only 30\% of the total shock luminosity in the model presented in Table \ref{Table1}.  In addition, we note that in our integrated spectra, only a portion of the precursor emission is captured within the aperture covering the SNR remnant proper, since the precursor emission region extends over the full area of our data cube.

In order to quantify the maximal effect of the precursor emission on our estimated depletion factors, we have compared the forbidden Fe line strengths with and without the effect of the (full) precursor emission. For individual lines arising from a given ionisation stage, the effect on the inferred depletion factor turns out to be almost the same, and is positive in all cases. That is to say that, in the theoretical models,  all iron lines are weakened with respect to the hydrogen recombination lines when the precursor is included. However, the effect is small, and it does not vary strongly from one ionisation stage to the next. To summarise, the following values should be added to the  depletion factors given in Table \ref{Table3} if the precursor is present in its full intensity: [Fe\,II] $\Delta \log(D) = 0.121$, [Fe\,III] $\Delta \log(D) = 0.050$, [Fe\,V] $\Delta \log(D) = 0.055$, [Fe\,VI] $\Delta \log(D) = 0.118$ and [Fe\,VII] $\Delta \log(D) = 0.149$. Note that applying these corrections would increase the requirement for grain destruction between the zone emitting in [Fe\,III] and the zone emitting in [Fe\,II].

\subsection{Effect of Multiple Shocks and Thermal Instabilities}
There is no doubt that the single shock model we have used is a severe oversimplification of reality. Both observation and theoretical work by others, and our discussion on the phase structure of the ISM and on the dynamical separation between the [\ion{Fe}{2}] and [\ion{Fe}{14}] emissions shows that a wide range of shock velocities exists in N49. Not only is there likely to be a large range in pre-shock density, but also thermal instabilities must be taken into account. 

As far as thermal instabilities are concerned, simple numerical or hydro 1-D and steady flow 2-D treatments by \citet{Chevalier82,Imamura84,Gaetz88} and \citet{Strickland95} show that in non-magnetic shocks the instability leads to periodic limit cycles caused by cooling and loss of pressure in the post-shock region with the formation of secondary shocks followed by the re-establishment of a hot pad of gas. This treatment was extended to magnetically supported shocks by \citet{Innes92}, who showed that a strong enough transverse magnetic field could suppress the formation of secondary shocks. Up to now the only fully 2-D treatment of thermally-unstable shocks is by \citet{Sutherland03}. They found that cooling thermally unstable shocks have enhanced cooling efficiency, due to their fractal structure, compared to homogeneous one- and two-dimensional models. The increased radiative efficiency is accompanied by a decrease in the conversion of kinetic to thermal energy as the additional degrees of freedom in the two-dimensional models allow kinetic energy to be redirected in other spatial directions, resulting in two-dimensional turbulence and the formation of many secondary shocks within the cooling structure (a ``shocklet box"!). The effect of this is to make the shock structure masquerade somewhat as if it was a mixture of steady-flow shocks of lower velocity.

In this respect, thermal instabilities act somewhat as multiple shocks driven by a common pressure but with varying pre-shock density, since this will also produce a fractal range of shock velocities. This problem has been studied in detail by \citet{Vancura92a} who found that a power-law velocity distribution of shock velocity fitted best to the multi-wavelength N49 data. For comparison with the \citet{Vancura92a} and \cite{Vancura92b} observations, we present our ``best fit'' shock model with their observations in Table \ref{Table4}. We have followed the lead of the original authors in choosing to normalise our fluxes with respect to \ion{C}{4}. However, we should note the halos of the LMC and the Milky Way are essentially opaque to the blue-shifted parts of the profiles of strong resonance lines such as \ion{C}{4}, \ion{Si}{4} and \ion{O}{6} Which may cut the observed fluxes by $\sim2$ for these lines. The simple one-velocity shock model does about as well as might be expected in reproducing the data.

\begin{table}[htbp]
\caption{The UV Lines intensities (C\,IV =100)  in N49  \citep{Vancura92a,Vancura92b} compared with our single velocity shock model described in the text.}\label{Table4}
\begin{center}
\begin{tabular}{lccc}
\\ \hline \hline 
{\bf UV lines:} & & &\\
Ion & $\lambda$(\AA) & Obs. & Model  \\
\hline 
 {O VI} & 1035 &  405 & 536 \\
 {N V} & 1240 &  $\leq 25$ & 32\\
 {Si IV} & 1397 & 19 & 19.7 \\
 {O IV]} & 1403 & 39 & 88.7 \\
 {C IV} & 1549 & 100 & 100 \\
 {He II} & 1640 & 60 & 34.3 \\
 {O III]} & 1663 &  39 & 64.3 \\
 {Si III]} & 1892 &  49 & 65.7 \\
 {C III]} & 1909 &  81 & 41.0 \\
 {C II]} & 2325 &  81 & 24.6 \\
 {[Ne IV]} & 2423 &  17 & 24.6 \\
 {[O II]} & 2470 &  27 & 11.3 \\
  {Mg II} & 2800 &  54 & 68.1 \\
 \hline
\end{tabular}
\end{center}
\end{table}

As far as the Fe depletion factors derived from our observations are concerned, we know that we cannot obtain accurate depletion factors from species such as \ion{Fe}{10} or \ion{Fe}{14} since these arise in the fastest shock propagating into the lowest density phase of the ISM. However, our result that the Fe-bearing grains are destroyed rapidly between the \ion{Fe}{3} and the \ion{Fe}{2} zones of the radiative shocks requires that, in the presence of a range of shock velocities, the theoretical flux ratio of the [\ion{Fe}{3}] and [\ion{Fe}{2}] for a given depletion factor should be fairly independent of both shock velocity and magnetic parameter. 

\begin{figure}[htb!]
\begin{centering}
\includegraphics[scale=0.5]{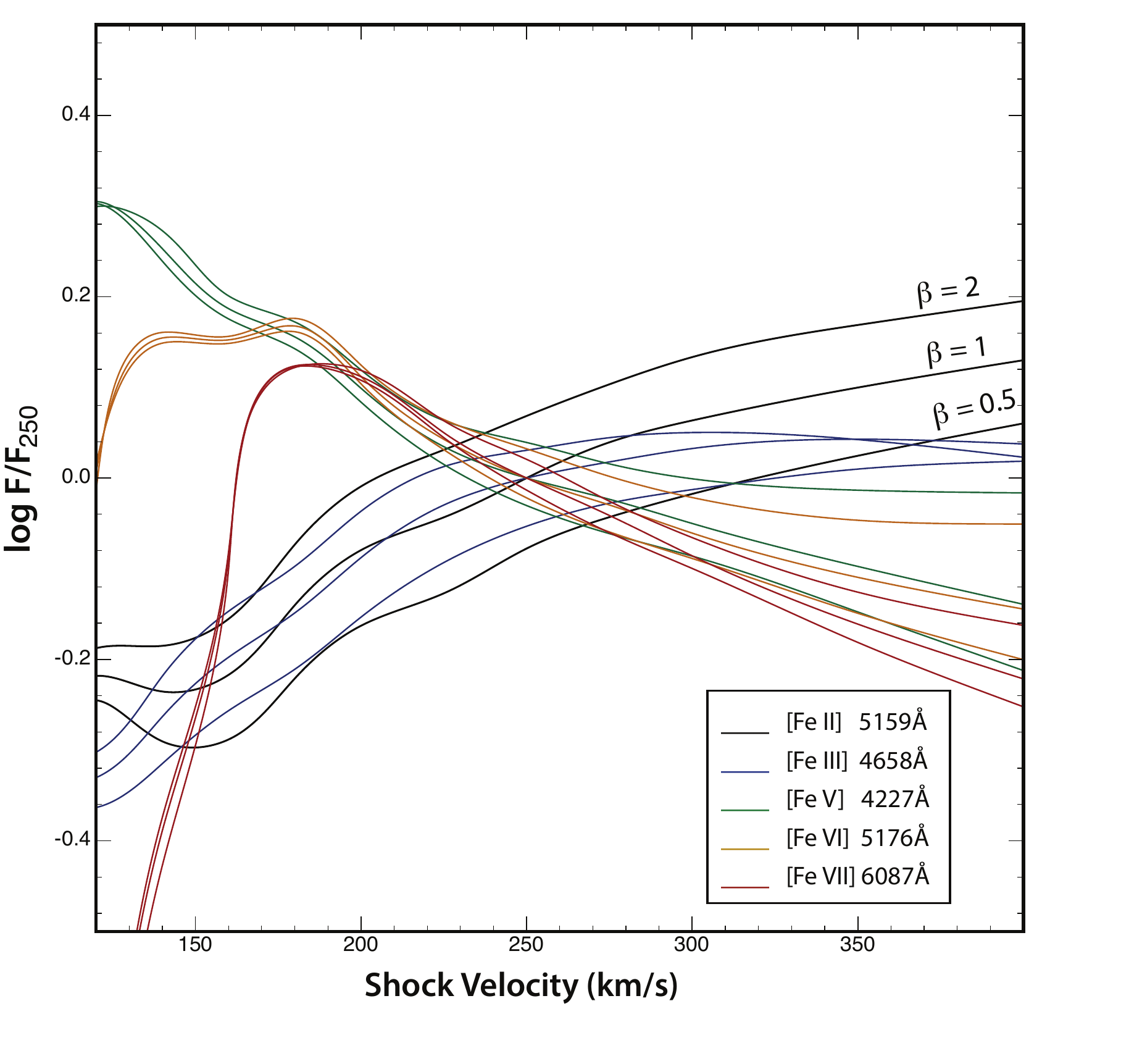}
\end{centering}
\caption{The variation of the relative fluxes of representative lines of Fe with respect to H$\beta$, as a function of shock velocity and magnetic $\beta$ normalised to a velocity of $\sim250$\,km\,s$^{-1}$ and a magnetic parameter $\beta = 1.0$. Note that the [\ion{Fe}{3}] and [\ion{Fe}{2}] lines vary in lockstep, to a good approximation. Thus the absolute errors on the derived depletion factors are $\lesssim \pm0.2$\,dex for shock velocities varying in the range $170 < v_s <  400$\,km\,s$^{-1}$, but the relative depletion factors are $\lesssim \pm0.1$\,dex over the same velocity range. }\label{fig10}
\end{figure}

This requirement is met by the shock models -- as we show in Figure \ref{fig10}. We find that the [\ion{Fe}{3}] and [\ion{Fe}{2}]  line intensities vary by less than $\sim \pm 0.2$\,dex in an absolute sense, and by $\sim \pm0.1$\,dex relatively over a velocity range $170 < v_s <  400$\,km\,s$^{-1}$. Note also that both the [\ion{Fe}{3}] and [\ion{Fe}{2}] lines increase in an absolute sense with velocity, as the collisional excitations in the photoionised region of the recombination zone becomes more important. By contrast, the higher stages of ionisation,  [\ion{Fe}{5}],  [\ion{Fe}{6}] and [\ion{Fe}{7}] become weaker with increasing velocity (provided that the shock is fast enough to produce these species in the first place). This variation results from the increasing relative emissivity of H$\beta$ with velocity, thanks to the increasing emissivity of the recombination region of the shock resulting from photoionisation by the shock UV radiation field.

In conclusion, if a mixture of shock velocities is present, this introduces an error of as much as $\sim0.2$\,dex in an absolute depletion factors we derive, and about  $\sim0.1$\,dex in the relative depletions of  \ion{Fe}{3} and \ion{Fe}{2}. This is much smaller than the $\sim0.8$\,dex inferred from the observations, so we can safely conclude that our result is robust against thermal instabilities, varying pre-shock ISM density and varying pre-shock magnetic $\beta$.

\section{Comparison with the Cygnus Loop SNR}
From UV, optical and IR data \citet{Sankrit14} found evidence for roughly 50\% destruction of dust grains in a $\sim150$\,km\,s$^{-1}$ shock, but in contrast to the our results, found that the the refractory elements C, Si and Fe were returned to the gas phase before the \ion{C}{3},  \ion{Si}{3} and  \ion{Fe}{3} line formation regions. In order to check this result, we ran a grid of shock models with pre-shock H density $n_{\mathrm H} = 10.0$\,cm$^{-3}$, with a local galactic concordance abundances,  but otherwise covering the same parameter range  as the LMC grid. We found that a very good agreement with the ``cusp region'' spectroscopy given in Table 2 of  \citet{Sankrit14} could be obtained with a model with $v_s = 200$\,km\,s$^{-1}$ and a magnetic $\beta=1.0$, with \emph{all} elements depleted by the amount corresponding to $\log D_{\mathrm Fe} = -0.7$. The radiative shock in the cusp region of the Cygnus Loop has not yet had time to become fully radiative \citet{Dopita83,Raymond88}. As a consequence, the final model fit was terminated at temperature of 7000K, as recommended by \citet{Sankrit14}.  This gives a peak electron density in the model of $n_e=150$cm$^{-3}$, in the mid range of that derived from the [S\,II] doublet,  $100 \lesssim n_e \lesssim250$cm$^{-3}$,

\begin{table}[htbp]
\caption{Lines intensities (H$\beta =100$)  in the Cygnus Loop `cusp' region by \citet{Sankrit14} compared with the 200 km\,s$^{-1}$ shock model described in the text.}\label{Table5}
\begin{center}
\begin{tabular}{lccc}
\\ \hline \hline 
{\bf UV lines:} & & &\\
Ion & $\lambda$(\AA) & Obs. & Model  \\
\hline 
 {N V} & 1241 &  84 & 164 \\
 {C II} & 1335 &  33 & 38\\
 {O IV],Si IV} & 1403 & 162 & 229 \\
 {N IV]} & 1486 &  92 & 28 \\
 {C IV} & 1549 & 183 & 301 \\
 {O III]} & 1665 & 145 & 127 \\
 {N III]} & 1750 &  31 & 27 \\
 {Si III]} & 1883 &  81 & 74 \\
 {C III]} & 1909 &  232 & 295 \\
 \hline 
{\bf IR lines:} & & &\\
Ion & $\lambda (\mu{\mathrm m})$ & Obs. & Model  \\
\hline 
{[Ne II]} & 12.81&  46.9 & 59.6 \\
{[Ne V]} & 14.33 &  6.1 & 4.0 \\
{[Ne III]} & 15.56 &  59.5 & 95.7\\
 {[Fe II]} & 17.93 &  9.6 & 6.0 \\
 {[Fe III]} & 22.92 & 8.7 & 8.9 \\
 {[Ne V]} & 24.32 &  6.1 & 4.9 \\
 {[Fe II]} & 24.50 & 1.9 & 1.1 \\
 {[O IV]} & 24.88 & 21.7 & 16.0 \\
 {[Fe II]} & 25.98 & 20.2 & 20.0 \\
 {[Fe III]} & 33.00 &  2.2 & 2.5 \\
 {[Si II]} & 34.80 & 62 & 275 \\
 {[Fe II]} & 35.34 & 5.1 & 4.8 \\
 \hline
\end{tabular}
\end{center}
\end{table}

In Table \ref{Table5} we give the observed and the predicted line intensities of the refractory elements C, Si and Fe lines given by this model. For comparison, and to constrain the shock velocity, these have been supplemented by the intensities of the N lines in the UV, the Ne lines in the IR and the O lines in both the UV and IR parts of the spectrum. The N\,V, N\,IV] and N\,III] lines in the UV serve to constrain the excitation conditions and hence the shock velocity. Likewise, the [Ne\,II], [Ne\,III] and [Ne\,V] lines in the IR serve in a similar manner. We use these lines to estimate the shock velocity of close to  $v_s = 200$\,km\,s$^{-1}$, rather higher than the  $v_s = 150-155$\,km\,s$^{-1}$ derived by \citet{Sankrit14}.  

Within the errors, the intensities of all Fe lines are well fit to a depletion of $\log D_{\mathrm Fe} = -0.7$. Overall the model and the observations are in excellent agreement. There is marginal evidence that C and Si may be slightly more depleted than given by the model. However, this effect might be in part due to line of sight interstellar absorption acting on the \ion{Si}{4}  \ion{C}{2} and \ion{C}{4} emission lines. Interstellar absorption might also help to reduce the discrepancy between the observed and predicted  \ion{N}{5} emission line intensities.

The depletion factor we derive here for Fe is considerably greater than the $-0.3 > \log D_{\mathrm Fe} > -0.4$ estimated by  \citet{Sankrit14}, and is more in line with the  $\log D_{\mathrm Fe} = -0.95$ we obtained from the [\ion{Fe}{3}] ion in N49.  What is more curious though, is the fact that in Cygnus, the depletion factor for the [\ion{Fe}{2}] lines appears to be about the same as for the [\ion{Fe}{3}] lines. We ascribe this difference to the fact that, unlike N49, the Cygnus loop shocks are not yet fully radiative. Thus the cool  [\ion{Fe}{2}] - emitting tail of the shock is missing. This is precisely the zone in which the N49 results imply the fastest grain destruction rates.

In conclusion, we find first that the UV plus the IR region of the spectrum is just as valuable as the visible in the investigation of grain destruction by shocks, that  the ISM depletion of Fe in Cygnus is only somewhat less than that of N49, and that the Cygnus Loop observations demonstrate the importance of the presence of the recombination tail of the radiative shocks in ensuring grain destruction.

\section{Discussion}
The major results of this paper are as follows:
\begin{enumerate}
\item{The shocks producing coronal line emission of iron; [\ion{Fe}{10}]$\lambda6087$ and [\ion{Fe}{14}]$\lambda5303$ are associated with the blast wave of N49, moving at higher velocity than the fully-radiative shocks which dominate the total emission, and which are produced in the interaction of the blast-wave with dense molecular clouds. The typical radiative shock has  $v_s = 200-250$\,km\,s$^{-1}$ and a pre-shock hydrogen density $n_{\rm H} = 80$\,cm$^{-3}$, while the blast wave is characterised by a shock velocity of  $v_s = 350-400$\,km\,s$^{-1}$ which implies a pre-shock density of  $n_{\rm H} \sim 25$\,cm$^{-3}$ -- a remarkably high density, but consistent with the observation that the ISM in the region of N49 is unusually dense in order for the SNR to have proceeded so rapidly into its radiative phase, and into its recombination phase in X-ray emission  \citep{Park12}.}
\item{The evidence of destruction of Fe--bearing grain materials adduced from the depletion analysis shows that, by the time these have passed through the recombination zone of the cloud shocks with $v_s = 200-250$\,km\,s$^{-1}$, approximately 70\% of the grains have been destroyed. However, in the cooling zone of the shocks less than 10\% of these grains have been destroyed. This result is supported by the observations of the `cusp' region of the Cygnus Loop. Here, Fe is depleted throughout the cooling region of the shock, but the recombination zone has only just started to form, due to the finite age of the shocks.}
\end{enumerate}

The issue of grain destruction deserves some further discussion. Naively, one might assume that thermal sputtering in the hot plasma would be the main destruction process. However, the dust grains have a high momentum on passage through the shock front, and collisions due to their relative motion through the post-shock medium come to dominate the sputtering. This is termed non-thermal sputtering \citep{Barlow78,McKee87,Jones94}. Thermal sputtering is confined to fast, non-radiative shocks \citep{Seab87}.

\citet{Flower95} considered non-thermal sputtering in the context of C-shocks in molecular clouds, and found that this grain destruction process may proceed efficiently in such shocks, even at rather low shock velocities. In radiative shocks, \citet{Seab83} considered non-thermal grain destruction in radiative shocks. They point out that He dominates in this process, and emphasise  the importance of gyro-acceleration or betatron acceleration \citep{Spitzer76} of the dust grains caused by compression of the transverse magnetic field during cooling. The smallest grains are efficiently decelerated, while the larger grains can reach relative velocities considerably in excess of the shock velocity. The maximum of velocity occurs close to the recombination region of the shock in their models. In this region, other forms of grain destruction are important, especially grain-grain collisions as considered in detail by \citet{Borkowski95}. Depending on the relative velocity, either sputtering occurs, or cratering with the production of many small fragments or else, at highest velocities, catastrophic fragmentation (shattering). \citet{Borkowski95} estimate that  assuming an initial MRN \citep{Mathis77} grain size distribution and an injection velocity of 200\,km\,s$^{-1}$, the sputtered and vaporised mass fraction returned to the gas phase is 64\%. This number is remarkably close to our estimated gas-phase fraction derived from the  [\ion{Fe}{2}] depletion; 70\%. However, the observational uncertainties allow a fairly generous error as discussed above and the actual fraction destroyed could lie anywhere between 30\% and 90\%.

More recently \citet{Slavin15} computed the grain destruction in evolving supernova remnants, allowing for the change in shock velocity over time, as the SNR is slowed by its interaction with increasing amounts of interstellar gas during its Sedov phase. They find that carbonaceous grains are more difficult to destroy than the siliceous grains. At a shock velocity of 250\,km\,s$^{-1}$,  they expect about 50\% of the siliceous grains to be destroyed, and most of what remains in the form of shattered small grains. This result is not in conflict with our observational result. If the Fe is contained in the form of iron magnesium silicates, then perhaps the  \citet{Slavin15} computations apply. If the Fe in the grains is mostly in the oxide form, then we cannot readily estimate the fraction destroyed. However, interestingly they find that doubling the explosion energy of the supernova approximately doubles the amount of dust mass destroyed. 

A central difficulty with the \citet{Slavin15} results arises because the SNR evolution and corresponding grain destruction has only been considered for a homogenous ISM. We have abundant evidence presented above, and in the papers referred to in the text that in N49, the blast wave is evolving in a low density medium, while the fully radiative shocks exist in the denser phases of the ISM. It is therefore difficult to quantify the amount of grain destruction to be expected from the theoretical computations.

We conclude that we have established a new and fairly robust method of analysing grain destruction in SNR shocks by detailed analysis of the forbidden Fe line spectrum in the optical over many stages of ionisation. Our observations of the depletion of iron in N49 provide a strong qualitative confirmation of the theory developed by \citet{Seab83}  and applied by \citet{Borkowski95}, and \citet{Slavin15} in that most of the grain destruction occurs in the recombination zone of the fast radiative shocks. However, there remain important uncertainties in both the theory of grain destruction, and in the interpretation of the observational results. The theoretical computations depend on the phase structure of the ISM, on the grain composition, on the explosion energy, and on the temporal evolution of the SNR, while the observational results have uncertainties due to the input atomic physics, and due to the dependence of the results on the phase structure of the ISM, the age of the shocks and on the role of thermally unstable cooling.

On the observational side, it would be very useful to apply integral field techniques at higher spatial resolution, for example data with MUSE, to investigate the spatial variability of grain destruction, and its detailed dependence upon the local shock velocity. On the theoretical side it would be valuable to extend the work of  \citet{Slavin15} to a more realistic fractal structure of the ISM.
 
\section*{Acknowledgements}
The authors wish to thank the anonymous referee for valuable and constructive comments which have enabled us to much improve the original manuscript. Dopita acknowledges the support of the Australian Research Council (ARC) through Discovery project DP130103925. Dopita would also like to thank the Deanship of Scientific Research (DSR), King AbdulAziz University for additional financial support as Distinguished Visiting Professor under the KAU Hi-Ci program. Seitenzahl was supported by Australian Research Council Laureate Grant FL0992131. Winkler acknowledges support of the NSF through grant AST-0908566. This research has made use of \textsc{SAOImage DS9} \citep{Joye03}, developed by Smithsonian Astrophysical Observatory, \textsc{Aladin sky atlas} developed at CDS, Strasbourg Observatory, France \citep{Bonnarel00, Boch14}, \textsc{Astropy}, a community-developed core Python package for Astronomy \citep{Astropy13}, \textsc{APLpy}, an open-source plotting package for Python hosted at \url{http://aplpy.github.com}, and of \textsc{Montage}, funded by the National Science Foundation under  Grant Number ACI-1440620, and was previously funded by the National Aeronautics and Space Administration's Earth Science Technology Office, Computation Technologies Project, under Cooperative Agreement Number NCC5-626 between NASA and the California Institute of Technology. The "Second Epoch Survey" of the southern sky was made by the Anglo-Australian Observatory (AAO) with the UK Schmidt Telescope. Plates from this survey have been digitized and compressed by the STScI. The digitized images are copyright (c) 1993-2000 by the Anglo-Australian Observatory Board.

\bibliographystyle{apj}

\end{document}